\documentclass[journal]{IEEEtran}

\usepackage{graphicx}
\usepackage{subfigure}
\usepackage{color,soul}
\usepackage{setspace}

\usepackage[cmex10]{amsmath}
\usepackage{amssymb}
\usepackage{amsthm}
\usepackage{MnSymbol}
\usepackage{float}
\usepackage{epstopdf}
\usepackage{longtable}
\usepackage{url}
\usepackage{cite}
\usepackage{bm}
\usepackage{multirow}
\usepackage{algorithmic}
\usepackage{enumerate}
\usepackage{amsthm}
\usepackage{tikz}

\def\x{{\mathbf x}}

\def\w{{\mathbf w}}
\def\y{{\mathbf y}}
\def\z{{\mathbf z}}

\def\N{{\cal N}}
\def\S{{\cal P}}
\def\Y{{\cal Y}}

\def\F{{\mathcal F}}

\def\R{{\mathbb{R}}}
\def\n!{{\!\!\!\!\!}}
\def\l{{\ell}}

\def\ll{{\lsem}}
\def\rr{{\rsem}}

\newcommand{\eqdef}{\mathrel{\overset{\makebox[0pt]{\mbox{\normalfont\tiny\sffamily def}}}{=}}}

\def\pgf{{M}}
\newcommand{\Dotp}[2]{\langle #1, #2 \rangle}

\newtheorem{thm}{Theorem}
\newtheorem{lem}{Lemma}

\begin{document}

\title{Multisensor CPHD filter}

\author{Santosh~Nannuru, 
				Stephane~Blouin, %
        Mark~Coates,~\IEEEmembership{Senior Member,~IEEE }
				and Michael~Rabbat %
\thanks{S. Nannuru, M. Coates and M. Rabbat are with the Department
of Electrical and Computer Engineering, McGill University, Montreal,
QC, Canada. S. Blouin is with DRDC Atlantic Research Centre, Halifax, 
Canada. e-mail: santosh.nannuru@mail.mcgill.ca, 
Stephane.Blouin@drdc-rddc.gc.ca, 
mark.coates@mcgill.ca, michael.rabbat@mcgill.ca. This research was
conducted under PWGSC contract W7707-145675/001/HAL supported by Defence
R\&D Canada.}}

\maketitle

\begin{abstract}
The single sensor \emph{probability hypothesis density} (PHD) and  
\emph{cardinalized probability hypothesis density} (CPHD) filters have 
been developed in the literature using the random finite set framework. 
The existing multisensor extensions of these filters have limitations 
such as sensor order dependence, numerical instability or high computational 
requirements. In this paper we derive update equations for the multisensor 
CPHD filter. The multisensor PHD filter is derived as a special case.
Exact implementation of the multisensor CPHD involves sums over all 
partitions of the measurements from different sensors and is thus intractable. 
We propose a computationally tractable approximation which combines a greedy 
measurement partitioning algorithm with the Gaussian mixture representation 
of the PHD. Our greedy approximation method allows the user to control the 
tradeoff between computational overhead and approximation accuracy.
\end{abstract}


\section{Introduction}
\label{sec:introduction}

In the multitarget tracking problem often the number of targets and
the number of observations detected by sensors are unknown and time
varying. Thus representing the targets and observations as vectors 
is inefficient. A finite set is a more suitable representation. This 
is the motivation for the random finite set 
framework~\cite{goodman1997,mahler2007B} which represents target 
states and observations as realizations of random finite sets. 
The implementation of the general multitarget Bayes filter for 
random finite sets is analytically and computationally 
infeasible~\cite{mahler2003}. 
Several approximations have been proposed which make 
suitable assumptions to derive tractable 
filters~\cite{mahler2003,mahler2007,vo2006,vo2007,vo2005}.

The majority of research based on random finite set theory has focused
on single sensor multitarget tracking. The \emph{probability
  hypothesis density} (PHD) filter~\cite{mahler2003} propagates, over
time, the probability hypothesis density function which is defined
over the single target state space. Improving on the PHD filter, the
\emph{cardinalized probability hypothesis density} (CPHD)
filter~\cite{mahler2007} propagates the distribution of the number of
targets (the cardinality) in addition to the PHD function. Various
implementations of the PHD and CPHD filter have been proposed,
including the Gaussian mixture implementation~\cite{vo2006,vo2007} and
the sequential Monte Carlo implementation~\cite{vo2005}. These
algorithms have been successfully applied to the problem of
multitarget tracking in the presence of clutter.

A general multisensor extension of the PHD filter was first derived for 
the case of two sensors by Mahler~\cite{mahler2009a,mahler2009b}. The 
filter equations were further generalized to include an arbitrary number 
of sensors by Delande et al.\ \cite{delande2010}. Because of their
combinatorial nature, the exact filter update 
equations of the general multisensor PHD filter are not computationally 
tractable except for a very few simple cases. Delande et al.\
\cite{delande2011a,delande2011b} derive simplifications to the 
filter update equations for the case when the fields of view of different sensors have 
limited overlap. This reduces the computational complexity to some extent, 
and a particle filter based implementation is presented in~\cite{delande2011b}. Jian et al.\ 
\cite{jian2013} suggest implementing the general multisensor PHD filter 
by repeated application of the two sensor PHD filter~\cite{mahler2009a}. 
The implementation details for realizing the general multisensor PHD filter 
in this manner are not made explicit, and the reported numerical simulations 
are restricted to the case of two sensors.

To avoid the combinatorial computational complexity of the general
multisensor PHD filter, some approximate multisensor filters have been
proposed in the literature. The \emph{iterated-corrector PHD filter} 
\cite{mahler2009b} processes the information from different sensors in 
a sequential manner. A single sensor PHD filter processes measurements 
from the first sensor. Using the output PHD function produced by this 
step as the predicted PHD function, another single sensor PHD filter 
processes measurements from the second sensor and so on. As a result, 
the final output strongly depends on the order in which sensors are 
processed~\cite{nagappa2011}. This dependence on the sensor order can 
be mitigated by employing the \emph{approximate product multisensor 
PHD and CPHD filters} proposed by Mahler~\cite{mahler2010}. Although 
the final results are independent of sensor order, Ouyang and 
Ji~\cite{ouyang2011} have reported that Monte Carlo implementation 
of the approximate product multisensor PHD filter is unstable and the 
problem worsens as the number of sensors increases. We have observed 
a similar instability in Gaussian mixture model-based implementations. 
Ouyang and Ji~\cite{ouyang2011} have proposed a heuristic fix to 
stabilise the Monte Carlo implementation but it is not analytically verified.
A comprehensive review of the different multisensor multitarget tracking algorithms 
based on random finite set theory can be found in~\cite[Ch.~10]{mahler2014B}.

In this paper we derive the update equations for the general multisensor 
CPHD filter. The derivation method is similar to that of the general 
multisensor PHD filter~\cite{mahler2009a,delande2010} with the additional 
propagation of the cardinality distribution. The multisensor CPHD
filter we derive has combinatorial complexity and an exact 
implementation is computationally infeasible. To overcome this
limitation we propose a two-step greedy approach based on a Gaussian
mixture model implementation. Each step can be realized using a trellis
structure constructed using the measurements from different sensors or 
measurement subsets for different Gaussian components.
The algorithm is applicable to both the general multisensor CPHD and 
the general multisensor PHD filters.

Other trellis based algorithms have been developed for target tracking. 
For single-sensor single-target tracking, the Viterbi algorithm is applied 
over a trellis of measurements constructed over time in~\cite{lascala1998}.
Each column of the trellis is a measurement scan at a different time step. 
The Viterbi algorithm is used to find the best path in the trellis 
corresponding to data associations over time. This approach has been 
extended to multitarget tracking in~\cite{pulford2006} for a fixed and
known number of targets. The nodes of the trellis correspond to different
data association hypotheses and the transition weights are based on 
measurement likelihoods. The Viterbi algorithm was also applied
in~\cite{wolf1989}, in conjunction with energy based transition weights, 
to identify the $K$-best non-intersecting paths over the measurement trellis 
when $K$ targets are present.

The form of the update equations in the general multisensor PHD/CPHD 
filters are similar to the update equations of the single sensor PHD/CPHD 
filters for extended targets~\cite{mahler2009c,orguner2011}. The similarity 
is in the sense that for extended targets the update equation requires 
partitioning of the single sensor measurement set which can be 
computationally demanding. Granstrom et al.\ \cite{granstrom2010} propose 
a Gaussian mixture model-based implementation of the PHD filter for extended
targets with reduced partitioning complexity. This is done by calculating 
the Mahalanobis distance between the measurements and grouping together 
measurements which are close to each other within a certain threshold.
Orguner et al.\ \cite{orguner2011} use a similar method to reduce computations 
in the Gaussian mixture model-based implementation of the CPHD filter for 
extended targets.

The rest of the paper is organized as follows: Section~\ref{sec:background}
provides a brief overview of random finite sets. Section~\ref{sec:problem_formulation}
formally poses the problem of multisensor multitarget tracking. In 
Section~\ref{sec:gcphd_filter} we summarize the prediction and update equations 
of the general multisensor CPHD filter. The derivation of the filter update 
equations is provided in the appendices. We present computationally tractable 
implementations of the general multisensor PHD and CPHD filters in 
Section~\ref{sec:apprx_implementation}. A performance comparison of the proposed 
filter with existing multisensor filters is conducted using numerical simulations 
in Section~\ref{sec:simulations}. We provide conclusions in Section~\ref{sec:conclusion}.

Portions of this work are presented in a conference paper~\cite{nannuru2014}; 
the present manuscript contains detailed derivations and proofs which were 
omitted from the conference paper, and the present manuscript also includes 
a more detailed description and evaluation of the proposed approximation 
and implementation of the general multisensor PHD and CPHD filters.

\section{Background on random finite sets}
\label{sec:background}

\subsection{Random finite sets}
\label{sec:rfs}

Random finite sets are set-valued random variables. The PHD and CPHD 
filters are derived using notions of random finite sets. This section 
provides a review of this background, introducing definitions and notation 
used in the derivations that follow. Detailed treatments of random finite 
sets and the related statistics in the context of multitarget tracking can 
be found in~\cite{goodman1997, mahler2007B, mahler2014B}.

A random finite set is completely specified using its probability density 
function if it exists. The probability density function of a random finite 
set modeling the multitarget state is also referred to as the multitarget 
density function in this paper. Let $Y$ be a realization of a random finite 
set $\Xi$ with elements from an underlying space $\Y$, i.e. 
$Y \subseteq \Y$. For the random finite set $\Xi$, denote its 
density function by $f_{\Xi}(Y)$. Let the cardinality distribution of the 
random finite set $\Xi$ be $p_{\Xi}(n)$ 
\begin{align}
p_{\Xi}(n) \eqdef \textrm{Prob}(|\Xi| = n) \,, \; n = 1,2,\dots \,,
\end{align}
where the notation $|\Xi|$ denotes the cardinality of set $\Xi$. The 
\emph{probability generating function} (PGF) of the cardinality distribution 
$p_{\Xi}(n)$ is defined as
\begin{align}
\pgf_{\Xi}(t) \eqdef \sum_{n=0}^{\infty} t^n p_{\Xi}(n).
\end{align}

A statistic of the random finite set, which is used by the PHD and CPHD filters, 
is the probability hypothesis density (PHD) function~\cite{mahler2007B}. For the 
random finite set $\Xi$ defined over an underlying space $\Y$, we denote 
its PHD by $D_{\Xi}(\y)$, $\y \in \Y$. Unlike the probability density
function $f_{\Xi}(Y)$ which is defined over the space of finite sets in 
$\Y$, the PHD function $D_{\Xi}(\y)$ is defined over the space 
$\Y$. Instead of propagating the complete density function, which can be 
computationally challenging, the PHD and CPHD filters propagate the low 
dimensional PHD function over time.

\subsection{IIDC random finite set}

An \emph{independent and identically distributed cluster} (IIDC) random finite 
set~\cite{mahler2007} is completely specified by its cardinality distribution 
and its spatial density function. Let $\Xi$ be an IIDC random finite set with 
cardinality distribution $p_{\Xi}(n)$ and the spatial density function $\zeta(\y)$.
The probability density function $f_{\Xi}(Y)$ and the PHD $D_{\Xi}(\y)$ of the 
random finite set $\Xi$ are given by the relations
\begin{align}
f_{\Xi}(Y) &= |Y|! \; p_{\Xi}(|Y|) \, \prod_{\y \in Y}{\zeta(\y)} \\
D_{\Xi}(\y) &= \zeta(\y) \, \mu \\
\mu &= E(|\Xi|) = \sum_{n=0}^{\infty} n \, p_{\Xi}(n).
\end{align}
Samples from an IIDC random finite set can be generated by first sampling a
cardinality $m$ from its cardinality distribution $p_{\Xi}(n)$ and then 
independently sampling $m$ points from its spatial density function $\zeta(\y)$.

The Poisson random finite set is an example of an IIDC random finite set 
where the cardinality distribution is assumed to be Poisson. The 
PHD filter~\cite{mahler2003} models the multitarget state as a realization of a 
Poisson random finite set and propagates its PHD function over time. The Poisson 
random finite set assumption can be undesirable because the variance 
of the Poisson distribution is equal to its mean, which implies that as the number of 
targets increases, the error in its estimation becomes larger. To overcome this 
problem the CPHD filter~\cite{mahler2007} models the multitarget state as a 
realization of an IIDC random finite set and propagates its PHD function and cardinality 
distribution over time. The additional cardinality information allows us to 
more accurately model the multitarget state.	

\section{Problem formulation}
\label{sec:problem_formulation}

We now specify the multisensor multitarget tracking problem. Let 
$\x_{k,i} \in \mathcal{X}$ be the state of the $i^{th}$ target at time
$k$. In most of the tracking literature $\mathcal{X}$ is chosen to be
the Euclidean space, $\mathcal{X} = \R^{n_{\x}}$, where $n_{\x}$ is the 
dimension of the single target state. If $n_k \geq 0$ targets are present 
at time $k$, the multitarget state can be represented by the finite set 
$X_k = \{\x_{k,1}, \ldots \x_{k,n_k} \}$, $X_k \subseteq \mathcal{X}$.
We assume that each single target state evolves according to the Markovian 
transition function $f_{k+1|k}(\x_{k+1,i}|\x_{k,i})$. New targets can 
arrive and existing targets can disappear at each time step. Let the 
survival probability of an existing target with state $\x$ at time $k$ 
be given by the function $p_{sv,k}(\x)$.

Multiple sensors make observations about the multiple targets present
within the monitored region. Assume that there are $s$ sensors, and 
conditional on the multitarget state, their observations are independent. 
Measurements $\z^{j}$ gathered by sensor $j$ lie in the space 
$\mathcal{Z}^{j}$, i.e., $\z^{j} \in \mathcal{Z}^{j}$. Let 
$Z^{j}_{k} = \{\z^{j}_{1,k}, \z^{j}_{2,k}, \dots, \z^{j}_{m_{j,k}}\}$, 
$Z^{j}_{k} \subseteq \mathcal{Z}^{j}$ be the set of measurements collected 
by the $j$-th sensor at time step $k$. The measurement set can be empty. 
We assume that each target generates at most one measurement per sensor 
at each time instant $k$. Each measurement is either associated with a 
target or is generated by the clutter process.
Define $Z^{1:s}_{k} = Z^{1}_{k} \cup Z^{2}_{k} \cup \dots \cup Z^{s}_{k}$ to be the
collection of measurement sets gathered by all sensors at time $k$.
The probability of detection of sensor $j$ at time $k$ is given by
$p^{j}_{d,k}(\x)$. The function $h_{j,k}(\z|\x)$ denotes the 
probability density (likelihood) that sensor $j$ makes a measurement 
$\z$ given that it detects a target with state $\x$. Denote the probability 
of a missed detection as $q^{j}_{d,k}(\x) = 1 - p^{j}_{d,k}(\x)$.

The objective of the multitarget tracking problem is to form an estimate
$\widehat{X}_k$ of the multitarget state at each time step $k$. This 
estimate is formed using all the measurements up until time $k$ obtained
from all the $s$ sensors which is denoted by 
$Z^{1:s}_{1:k} = \{Z^{1:s}_{1},Z^{1:s}_{2},\dots,Z^{1:s}_{k}\}$.
More generally, we would like to estimate the posterior multitarget state 
distribution $f_{k|k}(X_k|Z^{1:s}_{1:k})$.

\section{General multisensor CPHD filter}
\label{sec:gcphd_filter}

In this section we develop the CPHD filter equations when multiple sensors
are present. The derivation method is similar to the approach used to derive
the general multisensor PHD filter equations by Mahler~\cite{mahler2009a}
and Delande et al.\ \cite{delande2010}. 
Since the CPHD filter explicitly accounts for the cardinality distribution of the
multitarget state, the filter update equations are more involved. Specifically, 
the measurement set partitions are more explicitly listed when compared to the
PHD filter update equations~\cite{mahler2009a,delande2010}. The CPHD 
filter also requires additional propagation of the cardinality distribution.

We make the following modeling assumptions while deriving the multisensor 
CPHD filter equations
\newline
{\bf Assumption 1:} 
\begin{enumerate}[{a)}]
	\item Target birth at time $k+1$ is modelled using an IIDC random finite set. 
	\item The predicted multitarget distribution at time $k+1$ is IIDC.
	\item The sensor observation processes are independent conditional on the 
		multitarget state $X_{k+1}$, and the sensor clutter processes are IIDC.
\end{enumerate}

Before deriving the filter equations, we introduce some notation. 
Let $b_{k+1}(\x)$ be the PHD function and let $p_{b,k+1}(n)$ be 
the cardinality distribution of the birth process at time $k$.
For the $j$-th sensor let $c_{k+1,j}(\z)$ be the clutter spatial 
distribution and let $C_{k+1,j}(t)$ be the PGF of the clutter 
cardinality distribution at time $k+1$.
Let $D_{k+1|k}(\x)$ denote the predicted PHD function and let $r_{k+1|k}(\x)$ 
denote the normalized predicted PHD function at time $k+1$ (normalized so 
that it integrates to one). Let the PGF of the predicted cardinality 
distribution $p_{k+1|k}(n)$ be denoted by $\pgf_{k+1|k}(t)$. To keep the 
expressions and derivation compact we drop the time index and write
\begin{align}
c_{k+1,j}(\z) &\equiv c_{j}(\z), \; C_{k+1,j}(t) \equiv C_{j}(t), \;
	p^{j}_{d,k+1}(\x) \equiv p^{j}_{d}(\x) \nonumber \\
\pgf_{k+1|k}(t) &\equiv \pgf(t), \; q^{j}_{d,k+1}(\x) \equiv q^{j}_{d}(\x), \;
	p_{sv,k+1}(\x) \equiv p_{sv}(\x) \nonumber \\
h_{j,k+1}(\z|\x) &\equiv h_{j}(\z|\x), \; r_{k+1|k}(\x) \equiv r(\x), \; 
	m_{j,k+1} \equiv m_{j} \nonumber \\
p_{b,k+1}(n) &\equiv p_{b}(n) \,,
\end{align}
when the time is clear from the context.
Note that abbreviated notation is used only for convenience and the above 
quantities are in general functions of time. For functions $a(\y)$ and 
$b(\y)$, the notation $\Dotp{a}{b}$ is defined as 
$\Dotp{a}{b} = \int{a(\y) \, b(\y) \, d\y}$. In the following subsections 
we discuss the prediction and update steps of the general multisensor 
CPHD filter.

\subsection{CPHD prediction step}

Since sensor information is not required in the prediction step, the 
prediction step of the CPHD filter for the multisensor case is the 
same as that for the single sensor case.
Denote the posterior probability hypothesis density at time $k$ as 
$D_{k|k}(\x)$ and the posterior cardinality distribution as $p_{k|k}(n)$.
The predicted probability hypothesis density function at time $k+1$ is 
given by~\cite{mahler2007,vo2007}
\begin{align}
D_{k+1|k}(\x) &= b_{k+1}(\x) + \int \! {p_{sv}(\w)f_{k+1|k}(\x|\w)} D_{k|k}(\w) d\w \,,
\end{align}
where the integral is over the complete single target state space.
The predicted cardinality distribution at time $k+1$ is given 
by~\cite{mahler2007,vo2007}
\begin{align}
& p_{k+1|k}(n)  = \nonumber\\
&\, \sum_{j=0}^{n}p_b(n-j) 
		\sum_{l=j}^{\infty}{{l \choose j}	\frac{{\Dotp{D_{k|k}}{p_{sv}}}^j 
		{\Dotp{D_{k|k}}{1-p_{sv}}}^{l-j}}{{\Dotp{D_{k|k}}{1}}^l} p_{k|k}(l)} \,,
\end{align}
where $n$, $j$ and $l$ are non-negative integers.
The normalized predicted PHD function is given by
\begin{align}
r(\x) \equiv r_{k+1|k}(\x) &= \frac{D_{k+1|k}(\x)}{\mu_{k+1|k}}, \\
\text{where } \quad \mu_{k+1|k} &= \sum_{n=1}^{\infty}{n \, p_{k+1|k}(n)}.
\end{align}

\subsection{CPHD update step}

We use the notation $\ll 1 , s \rr$ to denote the set of integers from $1$ to $s$.
Let $W \subseteq Z^{1:s}_{k+1}$ such that for all $j \in \ll 1,s \rr$, 
$|W|_j \leq 1$ where $|W|_j = |\{ \z \in W : \z \in Z^{j}_{k+1}\}|$. 
Thus the subset $W$ can have at most one measurement from each sensor. Let 
$\mathcal{W}$ be the set of all such $W$. For any measurement subset $W$ 
we can uniquely associate with it a set of pair of indices $T_{W}$ defined as 
$T_{W} = \{(j,l): \z^{j}_{l} \in W\}$. For disjoint subsets $W_1, W_2, \dots W_{n}$, let 
$V = Z^{1:s}_{k+1} \setminus (\cup_{i=1}^{n}{W_{i}})$, so that $W_1, W_2, \dots W_{n}$ 
and $V$ partition $Z^{1:s}_{k+1}$. Think of the set $W_{i}$ as a collection of 
measurements made by different sensors, all of which are generated by the same 
target and the set $V$ as the collection of clutter measurements made by all 
the sensors. Let $P$ be a partition of $Z^{1:s}_{k+1}$, constructed using elements 
from the set $\mathcal{W}$ and a set $V$, given by
\begin{flalign}
P &= \{W_1, W_2, \dots W_{|P|-1}, V \}, \label{eq:partition_1} \\
\text{such that } & \bigcup_{i=1}^{|P|-1} W_{i} \cup V = Z^{1:s}_{k+1}, \label{eq:partition_2}&& \\
& W_i \cap W_j = \emptyset, \text{for any } W_i, W_j \in P, i \neq j \label{eq:partition_3} \\
& W_i \cap V = \emptyset, \text{for any } W_i \in P \label{eq:partition_4}
\end{flalign}
where $|P|$ denotes the number of elements in the partition $P$.

The partition $P$ groups the measurements in $Z^{1:s}_{k+1}$ into disjoint 
subsets where each subset is either generated by a target (the $W$ subsets) 
or generated by the clutter process (the $V$ subset). Let $|P|_{j}$ be the 
number of measurements made by sensor $j$ which are generated by the
targets. We have:
\begin{align}
|P|_{j} &= \sum_{i=1}^{|P|-1}{|W_i|_j}.
\end{align}
The number of measurements made by sensor $j$ which are classified as 
clutter in the partition $P$ is $(m_{j} - |P|_{j})$. Let $\S$ be the 
collection of all possible partitions $P$ of $Z^{1:s}_{k+1}$ constructed 
as above. A recursive expression for constructing the collection $\S$ is 
given in Appendix~\ref{app:recursive}.

Denote the $v^{th}$-order derivatives of the PGFs of the clutter 
cardinality distribution and the predicted cardinality distribution
as
\begin{align}
C^{(v)}_{j}(t) &= \frac{d^v C_{j}}{d t^v}(t) \,, \quad
\pgf^{(v)}(t) = \frac{d^v \pgf}{d t^v}(t).
\end{align}

We use $\gamma$ to denote the probability, under the predictive PHD, 
that a target is detected by no sensor, and we thus have:
\begin{equation}
\gamma \eqdef \int \! r(\x) \prod_{j=1}^{s}{q^{j}_{d}(\x)} d\x \,. \label{eq:gamma}
\end{equation}

For concise specification of the update equations, it is useful to
combine the terms associated with the PGF of the clutter cardinality
distribution for a partition $P$. Let us define the quantity
\begin{align}
\kappa_{P} \eqdef \prod_{j=1}^{s} C^{(m_{j}-|P|_{j})}_{j}(0). \label{eq:kappa_P}
\end{align}

For a set $W \in \mathcal{W}$ and the associated index set $T_{W}$ define 
the quantities
\begin{align}
&d_{W} \eqdef \frac{\displaystyle \int \!\! r(\x) \left( \prod_{(i,l) \in T_{W}}{\!\!\!\! p_{d}^{i}(\x)\,h_{i}(\z^{i}_{l}|\x)}\right) \, 
	\!\!\! \prod_{j:(j,*) \notin T_{W}}{\!\!\!\!\!\! q^{j}_{d}(\x)} d\x}
	{\displaystyle \prod_{(i,l) \in T_{W}}{c_{i}(\z^{i}_{l})}}, \label{eq:dW_cphd} \\
&\rho_{W}(\x) \eqdef \frac{\displaystyle \left( \prod_{(i,l) \in T_{W}}{\!\!\!\! p_{d}^{i}(\x)\,h_{i}(\z^{i}_{l}|\x)}\right) \, 
	\prod_{j:(j,*) \notin T_{W}}{\!\!\!\!\!\! q^{j}_{d}(\x)}}
	{\displaystyle \int \!\! r(\x) \left( \prod_{(i,l) \in T_{W}}{\!\!\!\! p_{d}^{i}(\x)\,h_{i}(\z^{i}_{l}|\x)}\right) \, 
	\prod_{j:(j,*) \notin T_{W}}{\!\!\!\!\!\! q^{j}_{d}(\x)} d\x}, \label{eq:rhoW_cphd}
\end{align}
where $(j,*)$ indicates any pair of indices of the form $(j,l)$.
The quantity $d_W$ can be interpreted as the ratio of the likelihood that 
the measurement subset $W$ was generated by the target process to the likelihood 
that the measurement subset $W$ was generated by the clutter process. 
The quantity $\rho_W(\x)$ can be interpreted as the normalized pseudolikelihood
contribution of the measurement subset $W$.

The updated probability hypothesis density function $D_{k+1|k+1}(\x)$
can be expressed as the product of the normalized predicted probability 
hypothesis density $r_{k+1|k}(\x)$ at time $k+1$ and a pseudolikelihood 
function.
The pseudolikelihood function can be expressed as a linear combination 
of functions (one function for each partition $P$) with associated
weights $\alpha_{P}$. The all-clutter partition $P = \{ V \}$ where
$V = Z^{1:s}_{k+1}$ has an associated weight $\alpha_{0}$. Define
\begin{align}	
\alpha_{0} &\eqdef \frac{\displaystyle \sum_{P \in \S} \left(\kappa_{P} \pgf^{(|P|)}(\gamma) \prod_{W \in P}{d_{W}}\right)}
	{\displaystyle \sum_{P \in \S} \left(\kappa_{P} \pgf^{(|P|-1)}(\gamma) \prod_{W \in
              P}{d_{W}}\right)} \,, \label{eq:alpha_equation_0} \\
\alpha_{P} &\eqdef \frac{\displaystyle \kappa_{P} \pgf^{(|P|-1)}(\gamma) \prod_{W \in P}{d_{W}}}
	{\displaystyle \sum_{Q \in \S} \left(\kappa_{Q} \pgf^{(|Q|-1)}(\gamma) \prod_{W \in
              Q}{d_{W}}\right)} \,. \label{eq:alpha_equation_P}
\end{align}
Note that the expression $W \in P$ only includes $W \in \mathcal{W}$ and 
does not include the component $V \in \mathcal{V}$ of $P$.
For the all clutter partition $P = \{ V \}$ there are no elements in the partition of the type $W$
and we use the convention $\prod_{W \in P}() = 1$ whenever $P = \{ V \}$.
Similary we use the convention $\sum_{W \in P}() = 0$ whenever $P = \{ V \}$.

\begin{thm} \label{thm:phd_update}
Under the conditions of Assumption 1, the general multisensor CPHD
filter update equation for the probability hypothesis density is
\begin{align}
\frac{D_{k+1|k+1}(\x)}{r_{k+1|k}(\x)} &= \alpha_{0} \, \prod_{j=1}^{s}{q^{j}_{d}(\x)}
	+ \sum_{P \in \S} \alpha_{P} \, \left(\sum_{W \in P} \rho_{W}(\x) \right)\, \label{eq:phd_update_gcphd}
\end{align}
and the update equation for the posterior cardinality distribution is
\begin{align}
\frac{p_{k+1|k+1}(n)}{p_{k+1|k}(n)} &= \frac{\displaystyle \sum_{\substack{P \in \S \\ |P| \leq n+1}} 
	\left(\kappa_{P} \frac{n!}{(n-|P| + 1)!} \gamma^{n-|P| + 1} \prod_{W \in P}{d_{W}}\right)}
	{\displaystyle \sum_{P \in \S} \left(\kappa_{P} \pgf^{(|P|-1)}(\gamma)  
	\prod_{W \in P}{d_{W}}\right)}\,, \label{eq:card_update}
\end{align}
where the quantities $\alpha_{0}$, $\alpha_{P}$, $\rho_{W}(\x)$ and $d_{W}$
are given in~\eqref{eq:alpha_equation_0}, \eqref{eq:alpha_equation_P},
\eqref{eq:rhoW_cphd} and~\eqref{eq:dW_cphd}, respectively.
\end{thm}
The proof of Theorem~\ref{thm:phd_update} is provided in Appendix~\ref{app:phd_update}.
It requires the concepts of functional derivatives, probability generating functionals 
and the multitarget Bayes filter which are revised in Appendices~\ref{app:functional_derivatives},
\ref{app:pgfl} and~\ref{app:multitarget_bayes} respectively. The proof depends on an 
intermediate result, Lemma~\ref{lem:F_derivative}, which is proved in 
Appendix~\ref{app:F_derivative}.

\subsection{General multisensor PHD filter as a special case}
In this section we show that the general multisensor PHD filter can be
obtained as a special case of the general multisensor CPHD filter when
the following assumptions are made.
\newline
{\bf Assumption 2:}
\begin{enumerate}[{a)}]
	\item Target birth at time $k+1$ is modelled using a Poisson random finite set. 
	\item The predicted multitarget distribution at time $k+1$ is Poisson.
	\item The sensor observation processes are independent conditional on the 
		multitarget state $X_{k+1}$, and the sensor clutter processes are Poisson.
\end{enumerate}

Since the multitarget state distribution is modelled as Poisson it suffices
to propagate the PHD function over time. Let the rate of the Poisson clutter 
process be $\lambda_{j}$ and let $c_{j}(\z)$ be the clutter spatial distribution 
for the $j^{th}$ sensor. Let $\mu_{k+1|k}$ be the mean predicted cardinality 
at time $k+1$. Using the Poisson assumptions for the predicted multitarget 
distribution and the sensor clutter processes we have
\begin{align}
\pgf^{(v)}(\gamma) &= \mu_{k+1|k}^{v} \; e^{\mu_{k+1|k}(\gamma - 1)} \\
C^{(v)}_{j}(0) &= \lambda_{j}^{v} \; e^{-\lambda_{j}}.
\end{align}
Using these in~\eqref{eq:alpha_equation_0} we have the simplification 
$\alpha_{0} = \mu_{k+1|k} $.
We can also simplify the term $\kappa_{P} \pgf^{(|P|-1)}(\gamma)$ as
\begin{align}
\kappa_{P} & \pgf^{(|P|-1)}(\gamma) = \left( \prod_{j=1}^{s} C^{(m_{j}-|P|_{j})}_{j}(0) \right) 
	\, \pgf^{(|P|-1)}(\gamma) \\
&= \left( \prod_{j=1}^{s} \lambda_{j}^{m_{j}-|P|_{j}} \, e^{-\lambda_{j}} \right) \, 
	\mu_{k+1|k}^{|P|-1} \; e^{\mu_{k+1|k}(\gamma - 1)} \\
&= (e^{\mu_{k+1|k}(\gamma - 1) - \sum_{j=1}^{s}\lambda_{j}}) 
	\left( \prod_{j=1}^{s} \lambda_{j}^{m_{j}} \right)
	\frac{\mu_{k+1|k}^{|P|-1}}{\left( \prod_{j=1}^{s} \lambda_{j}^{|P|_{j}} \right)}.
\end{align}
Since the expression $\kappa_{P} \pgf^{(|P|-1)}(\gamma)$ appears in both the numerator 
and the denominator of the term $\alpha_{P}$ in the PHD update expression, we can 
ignore the portion that is independent of $P$. Hence we have
\begin{align}
\kappa_{P} \pgf^{(|P|-1)}(\gamma) & \propto \mu_{k+1|k}^{|P|-1} 
	\displaystyle \prod_{j=1}^{s} \lambda_{j}^{-|P|_{j}}.
\label{eq:psi_P_phd}
\end{align}
From~\eqref{eq:dW_cphd} and~\eqref{eq:psi_P_phd}, we can write
\begin{align}
\kappa_{P} \pgf^{(|P|-1)}(\gamma) \prod_{W \in P}{d_{W}} & \propto
	\prod_{W \in P}{\tilde{d}_W}
\end{align}
where $\tilde{d}_W$ is defined as
\begin{align}
& \tilde{d}_W \eqdef  \nonumber \\
& \frac{\displaystyle \mu_{k+1|k} \int \!\! r(\x)
	\left( \prod_{(i,l) \in T_{W}}{\!\!\!\! p_{d}^{i}(\x)\,h_{i}(\z^{i}_{l}|\x)}\right) \, 
	\prod_{j:(j,*) \notin T_{W}}{\!\!\!\!\!\!  q^{j}_{d}(\x)} d\x}
	{\displaystyle \prod_{(i,l) \in T_{W}}{\lambda_{i} \, c_{i}(\z^{i}_{l})}} \cdot
\end{align}
The PHD update equation then reduces to
\begin{align}
&\frac{D_{k+1|k+1}(\x)}{r_{k+1|k}(\x)} = \nonumber \\ 
& \qquad \mu_{k+1|k} \, \prod_{j=1}^{s}{q^{j}_{d}(\x)}
	+ \sum_{P \in \S} \frac{\displaystyle \left( \prod_{W \in P}{\tilde{d}_{W}} \right) \, 
	\displaystyle \sum_{W \in P} \rho_{W}(\x)}
	{\displaystyle \sum_{P \in \S} \left(\prod_{W \in P}{\tilde{d}_{W}}\right)} \cdot
	\label{eq:phd_update_gphd}
\end{align}
The above equation is equivalent to the general multisensor PHD update 
equation given in~\cite{mahler2009a,delande2010}.

\section{Implementations of the general multisensor CPHD and PHD filters}
\label{sec:apprx_implementation}

In the previous section we derived update equations for the general 
multisensor CPHD filter which propagate the PHD function and cardinality 
distribution over time. Analytic propagation of these quantities is 
difficult in general without imposing further conditions. In the 
next subsection we develop a Gaussian mixture-based implementation 
of the filter update equations. Although the Gaussian mixture 
implementation is analytically tractable, it is computationally 
intractable. In Section~\ref{sec:measurement_subsets} 
and~\ref{sec:grouping} we propose greedy algorithms to drastically 
reduce computations and develop computationally tractable approximate 
implementations for the general multisensor CPHD and PHD filters.

\subsection{Gaussian mixture implementation}
\label{sec:gm_implementation}

We make the following assumptions to obtain closed form updates for 
equations~\eqref{eq:phd_update_gcphd} and~\eqref{eq:card_update}
\newline
{\bf Assumption 3:}
\begin{enumerate}[{a)}]
\item The probability of detection for each sensor is constant throughout 
	the single target state space; i.e., $p^{j}_{d}(\x) = p^{j}_{d}$, for all $\x$.
\item The predicted PHD is a mixture of weighted Gaussian densities.
\item The single sensor observations are linear functions of a single target 
	state corrupted by zero-mean Gaussian noise.
\item The predicted cardinality distribution has finite support; i.e., there 
	exists a positive integer $n_{0} < \infty$ such that $p_{k+1|k}(n) = 0$, 
	for all $n > n_{0}$.
\end{enumerate}
From the above assumptions we can express the normalized predicted PHD as
a Gaussian mixture model
\begin{align}
r(\x) &= \sum_{i=1}^{J_{k+1|k}}{w_{k+1|k}^{(i)} \, \N_{(i)}(\x)}
\end{align}
where $w_{k+1|k}^{(i)}$ are non-negative weights satisfying
$\sum_{i=1}^{J_{k+1|k}}{w_{k+1|k}^{(i)}} = 1$; and
$\N_{(i)}(\x) \eqdef	\N(\x;m_{k+1|k}^{(i)},\Sigma_{k+1|k}^{(i)})$ is the
Gaussian density function with mean $m_{k+1|k}^{(i)}$ and covariance matrix 
$\Sigma_{k+1|k}^{(i)}$. 
If $H_{j}$ is the observation matrix for sensor $j$ then its likelihood 
function can be expressed as $h_{j}(\z|\x) = \N(\z;H_{j}\x,\Sigma_{j})$.
Then under the conditions of Assumption 3, the posterior PHD at time $k+1$
can be expressed as a weighted mixture of Gaussian densities and the 
posterior cardinality distribution has a finite support.

Since the probability of detection is constant we have $\gamma = \prod_{j=1}^{s}{q^{j}_{d}}$.
For each partition $P$ the quantities $\pgf^{(|P|-1)}$ and $\pgf^{(|P|)}$
can be easily calculated since the predicted cardinality distribution has 
finite support. The integration in the numerator of~\eqref{eq:dW_cphd} is 
analytically solvable under Assumption 3 and using properties of Gaussian 
density functions~\cite{vo2006}. Hence $d_{W}$ can be analytically evaluated. 
From these quantities we can calculate $\alpha_{0}$ and $\alpha_{P}$ 
from~\eqref{eq:alpha_equation_0} and~\eqref{eq:alpha_equation_P}. For each 
measurement set $W$ we can express the product $r(\x) \rho_{W}(\x)$ as a sum 
of weighted Gaussian densities using the properties of Gaussian density 
functions~\cite{vo2006}. 
Thus from the update equation~\eqref{eq:phd_update_gcphd} the posterior PHD can 
be expressed as a mixture of Gaussian densities. Since the predicted cardinality 
distribution has finite support, from~\eqref{eq:card_update}, the posterior 
cardinality distribution also has finite support. Similarly, under appropriate 
linear Gaussian assumptions, the posterior PHD in~\eqref{eq:phd_update_gphd} 
can be expressed as a mixture of Gaussian densities.

The conditions of Assumption 3 allow us to analytically propagate 
the PHD and cardinality distribution but the propagation is still 
numerically infeasible. The combinatorial nature of the update step can 
be seen from~\eqref{eq:phd_update_gcphd}, \eqref{eq:card_update} 
and~\eqref{eq:phd_update_gphd}. Specifically, the exact implementation 
of the general multisensor CPHD and PHD filters would require evaluation 
of all the permissible partitions (i.e. all $P \in \S$) that could be 
constructed from all possible measurement subsets. The number of such 
partitions is prohibitively large and a direct implementation is 
infeasible. We now discuss an approximation of the update step to 
overcome this limitation.

The key idea of the approximate implementation is to identify elements of 
the collection $\S$ which make a significant contribution to the update 
expressions. We propose the following two-step greedy approximation to 
achieve this within the Gaussian mixture framework. The first approximation 
step is to select a few measurement subsets $W$ for each Gaussian component. 
These subsets are identified by evaluating a score function which quantifies the
likelihood that the subset was generated by that Gaussian component.
The second approximation step is to greedily construct partitions of these subsets 
which are significant for the update step. The following subsections
explain these two steps in detail.

\subsection{Selecting the best measurement subsets}
\label{sec:measurement_subsets}

A measurement subset is any subset of the measurement set $Z^{1:s}_{k+1}$ 
such that it contains at most one measurement per sensor. The total number 
of measurement subsets that can be constructed when the $j^{th}$ sensor 
records $m_{j}$ measurements is $\displaystyle \prod_{j=1}^{s}{(m_{j}+1)}$. 
When there are many targets present and/or the clutter rate is high this 
number can be very large. Since the size of the collection $\S$ depends 
on the number of measurement subsets, to develop a tractable implementation 
of the update step it is necessary to limit the number of measurement 
subsets. Instead of enumerating all possible measurement subsets, they are greedily 
and sequentially constructed and only a few are retained based on the 
scores associated with them.

Consider the measurement subset $W$ and the associated set $T_{W}$ as defined 
earlier. For the $i^{th}$ Gaussian component and the measurement subset 
$W$ we can associate a score function $\beta^{(i)}(W)$ defined as 
\begin{align}
\beta^{(i)}(W) &\eqdef \frac{\displaystyle \int \!\! \N_{(i)}(\x)
	\left( \prod_{(j,l) \in T_{W}}{\!\!\!\! p_{d}^{j}\,h_{j}(\z^{j}_{l}|\x)}\right)
	\left( \prod_{j:(j,*) \notin T_{W}}{\!\!\!\!\!\!  q^{j}_{d}} \right) d\x}
	{\displaystyle \prod_{(j,l) \in T_{W}}{c_{j}(\z^{j}_{l})}} \cdot
	\label{eq:beta_W}
\end{align}
The above score function is obtained by splitting the $d_{W}$ term 
in~\eqref{eq:dW_cphd} for each Gaussian component. Intuitively, this 
score can be interpreted as the ratio of the likelihood that the 
measurement subset $W$ was generated by the single target represented by 
the $i^{th}$ Gaussian component to the likelihood that the measurement 
subset $W$ was generated by the clutter process. The score $\beta^{(i)}(W)$ 
can be analytically calculated since the integral is solvable under 
Assumption 3 and using properties of Gaussian densities~\cite{vo2005}. 
The score is high when the elements of the set $W$ truly are the measurements 
caused by the target associated with the $i^{th}$ Gaussian component. We use 
$\beta^{(i)}(W)$ to rank measurement subsets for each Gaussian component 
and retain only a fraction of them with the highest scores.

\begin{figure}
\centering
\begin{tikzpicture}
	\foreach \x in {1,2,3,4,5}{
			\node at (1.5*\x-1.5,6) {$\x$};
	}
	\node at (3,6.5) {Sensor $(j)$};
	\foreach \x in {1,2,3,4,5}{
			\node[draw,circle] at (1.5*\x - 1.5,5) {$\z_{\emptyset}^{\x}$};
	}
	\foreach \y in {1,2,3}{
			\node[draw,circle] at (0,5-\y) {$\z_{\y}^{1}$};
	}
	\foreach \y in {1,2,3}{
			\node[draw,circle] at (1.5,5-\y) {$\z_{\y}^{2}$};
	}
	\foreach \y in {1,2}{
			\node[draw,circle] at (3,5-\y) {$\z_{\y}^{3}$};
	}
	\foreach \y in {1,2,3}{
			\node[draw,circle] at (4.5,5-\y) {$\z_{\y}^{4}$};
	}
	\foreach \y in {1}{
			\node[draw,circle] at (6,5-\y) {$\z_{\y}^{5}$};
	}	
	
	\draw[blue, ultra thick] (0.2,4.8) -- (1.5,3.7);
	\draw[blue, ultra thick] (1.5,3.7) -- (2.8,4);
		
	\draw[blue, thick] (0.2,2.8) -- (1.5,1.7);
	\draw[blue, thick] (1.5,1.7) -- (2.8,2.8);
	
	\draw[blue, thick] (0.2,3.2) -- (1.4,5.3);
	\draw[blue, thick] (1.4,5.3) -- (2.8,5);
	
	\foreach \y in {0,1,2,3}{
			\draw[red, thick,dashed] (3.2,4) -- (4.3,5-\y);
	}
\end{tikzpicture}
\caption{Trellis diagram for constructing measurement subsets for each
	Gaussian component. Solid blue lines represent measurement subsets 
	retained after processing sensors 1 to 3. Dashed red lines correspond 
	to extensions of retained measurement subsets when processing 
	measurements from sensor $4$.}
\label{fig:trellis_subsets}
\end{figure}
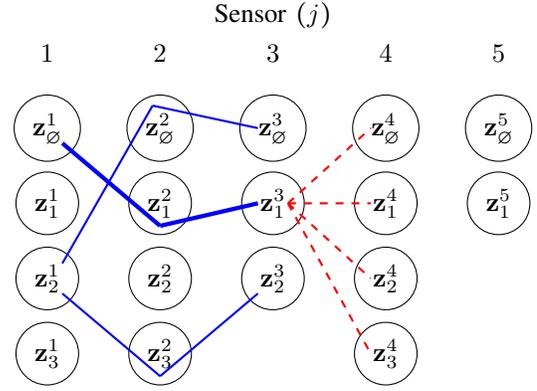

For each Gaussian component, we select the measurement subsets by
randomly ordering the sensors and incrementally incorporating 
information from each sensor in turn. We retain a maximum of
$W_{\textrm{max}}$ subsets at each step. Figure~\ref{fig:trellis_subsets} 
provides a graphical representation of the algorithm in the form of 
a trellis diagram. Each column of the trellis corresponds to observations 
from one of the sensors. The sensor number is indicated at the top of each 
column. The nodes of the trellis correspond to the sensor observations 
$(\z_{1}^{1}, \z_{2}^{1}, \dots, \, \z_{1}^{2}, \z_{2}^{2}, \dots)$ or 
the no detection case $(\z_{\emptyset}^{1}, \z_{\emptyset}^{2}, \dots)$.

The process of sequential construction of measurement subsets can 
be demonstrated using an example as follows. The solid lines in 
Figure~\ref{fig:trellis_subsets} represent partial measurement subsets 
retained after processing observations from sensors 1 to 3. Now 
consider the measurement subset indicated by the thick solid line. 
It corresponds to the measurement subset $\{\z_{1}^{2}, \z_{1}^{3}\}$.
When the sensor $4$ measurements are processed, this measurement subset 
is extended for each node of sensor $4$ as represented by the dashed 
lines. The scores $\beta^{(i)}(W)$ are calculated for these new 
measurement subsets using the expression in \eqref{eq:beta_W} but 
limited to only the first $4$ sensors. This is done for each existing 
measurement subset in the sensor-measurement space and 
$W_{\textrm{max}}$ measurement subsets with highest scores are 
retained and considered at the next sensor. Although the process of 
constructing measurement subsets is dependent on the order in which 
sensors are processed, we observe from simulations that it has no 
significant effect on filter performance. Once the subsets have been 
selected, the ordering has no further effect in the update process.

\subsection{Constructing partitions}
\label{sec:grouping}

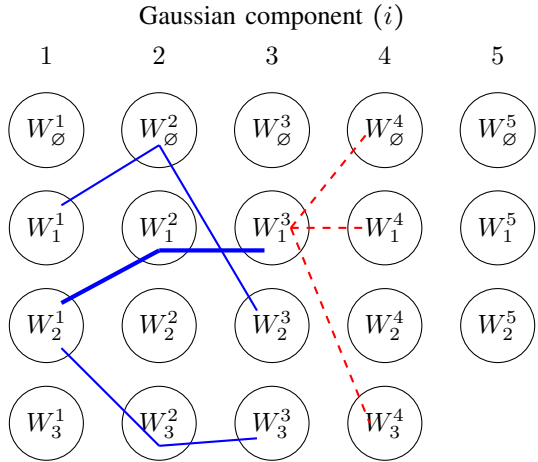
\begin{figure}
\centering
\begin{tikzpicture}
	\foreach \x in {1,2,3,4,5}{
			\node at (1.5*\x-1.5,6) {$\x$};
	}
	\node at (3,6.5) {Gaussian component $(i)$};
	\foreach \x in {1,2,3,4,5}{
			\node[draw,circle] at (1.5*\x - 1.5,5) {$W_{\emptyset}^{\x}$};
	}
	\foreach \y in {1,2,3}{
			\node[draw,circle] at (0,5-1.3*\y) {$W_{\y}^{1}$};
	}
	\foreach \y in {1,2,3}{
			\node[draw,circle] at (1.5,5-1.3*\y) {$W_{\y}^{2}$};
	}
	\foreach \y in {1,2,3}{
			\node[draw,circle] at (3,5-1.3*\y) {$W_{\y}^{3}$};
	}
	\foreach \y in {1,2,3}{
			\node[draw,circle] at (4.5,5-1.3*\y) {$W_{\y}^{4}$};
	}
	\foreach \y in {1,2}{
			\node[draw,circle] at (6,5-1.3*\y) {$W_{\y}^{5}$};
	}	
	
	\draw[blue, ultra thick] (0.2,2.7) -- (1.5,3.4);
	\draw[blue, ultra thick] (1.5,3.4) -- (2.9,3.4);
		
	\draw[blue, thick] (0.2,2.1) -- (1.5,0.8);
	\draw[blue, thick] (1.5,0.8) -- (2.8,0.9);
	
	\draw[blue, thick] (0.2,4) -- (1.5,4.8);
	\draw[blue, thick] (1.5,4.8) -- (2.8,2.6);
	
	\foreach \y in {0,1,3}{
			\draw[red, thick,dashed] (3.25,3.7) -- (4.3,5-1.3*\y);
	}
\end{tikzpicture}
\caption{Trellis diagram for constructing partitions.
	Solid blue lines represent partitions retained after processing Gaussian 
	components 1 to 3. Red dashed lines correspond to partition extension when 
	incorporating measurement subsets corresponding to the $4^{th}$ Gaussian 
	component.}
\label{fig:trellis_groups}
\end{figure}

The algorithm to construct partitions from subsets is similar to the above 
algorithm used to identify the best measurement subsets. Since the $V$ 
component of a partition is unique given the $W$ components, it is sufficient 
to identify the $W$ components to uniquely specify a partition $P$. A graphical 
representation of the algorithm is provided in Figure~\ref{fig:trellis_groups}. 
Each column of this trellis corresponds to the set of measurement subsets 
$\{W^{i}_{1}, W^{i}_{2}, \dots \}$ identified by the $i^{th}$ Gaussian 
component. The component number $(i)$ is indicated at the top of each column. 
The node $W_{\emptyset}^{i}$ represents the empty measurement subset 
$W_{\emptyset}^{i} = \emptyset$ which is always included for each component 
and it corresponds to the event that the Gaussian component was not detected 
by any of the sensors. With each valid partition $P$ we associate the score
$d_{P} = \displaystyle \prod_{W \in P}{d_{W}}$ 
with $d_{\emptyset} = 1$.

We greedily identify partitions of subsets by incrementally incorporating 
measurement subsets from the different components. For example the solid
lines in Figure~\ref{fig:trellis_groups} correspond to the partitions 
that have been retained after processing components number 1 to 3. The 
existing partitions are expanded using the measurement subsets from the 
$4^{th}$ component as indicated by the dashed lines. Some extensions are 
not included as they do not lead to a valid partition. Since the empty measurement 
subset $W_{\emptyset}^{i}$ is always included in the trellis, a partition can 
always be found. We process the 
Gaussian components in decreasing order of their associated weights. 
After processing each component, we retain a maximum of $P_{\textrm{max}}$ 
partitions corresponding to the ones with highest $d_{P}$. These selected 
partitions of measurement subsets are used in the update 
equations~\eqref{eq:phd_update_gcphd}, \eqref{eq:card_update} 
and~\eqref{eq:phd_update_gphd} to compute the posterior PHDs and 
cardinality distribution.
In our current implementation we select measurement subsets and construct 
measurement partitions using only the prior PHD information. Future research can
focus on enhancing this construction procedure by including the current 
measurements along with the prior PHD.

For the general multisensor PHD filter a slightly more accurate implementation
can be used and is described as follows. After the first approximate step of identifying measurement
subsets for each Gaussian component, instead of the approximate partition 
construction discussed in this section, we can find all possible partitions 
from the given collection of measurement subsets. This problem of finding 
all partitions can be mapped to the exact cover problem~\cite{michael1979}. 
An efficient algorithm called Dancing Links has been suggested by 
Knuth~\cite{knuth2000} for solving this problem. This implementation can be 
used when the number of sensors and measurement subsets are small.

\section{Numerical simulations}
\label{sec:simulations}

In this section we compare different multisensor multitarget tracking
algorithms developed using the random finite set theory. Specifically
we compare the following filters: 
iterated-corrector PHD (IC-PHD~\cite{mahler2009a}), 
iterated-corrector CPHD (IC-CPHD), general multisensor PHD (G-PHD) 
and the general multisensor CPHD (G-CPHD) filter derived in this paper. 
Models used to simulate multitarget motion and multisensor observations 
are discussed in detail in the next subsections. The simulated observations 
are used by different algorithms to perform multitarget tracking.
All the simulations were conducted using MATLAB~\footnote{The MATLAB 
code is available at http://networks.ece.mcgill.ca/software}.

\subsection{Target dynamics}

\begin{figure}[h]
	\centering
		\subfigure[Target tracks]{
			\includegraphics[width=0.46\textwidth]{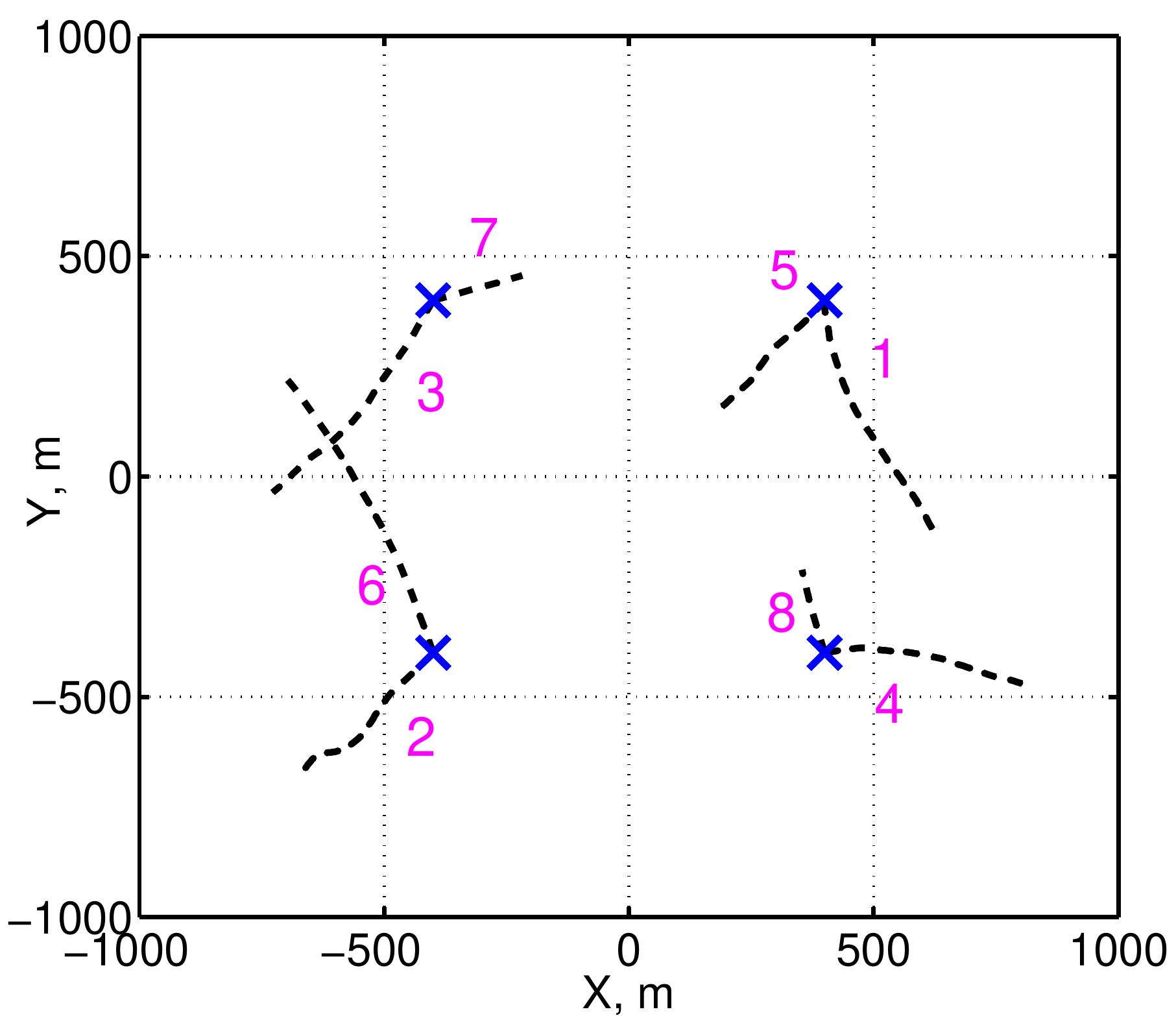}
			\label{fig:tracks_8t}}

		\subfigure[Cardinality]{
			\includegraphics[width=0.45\textwidth]{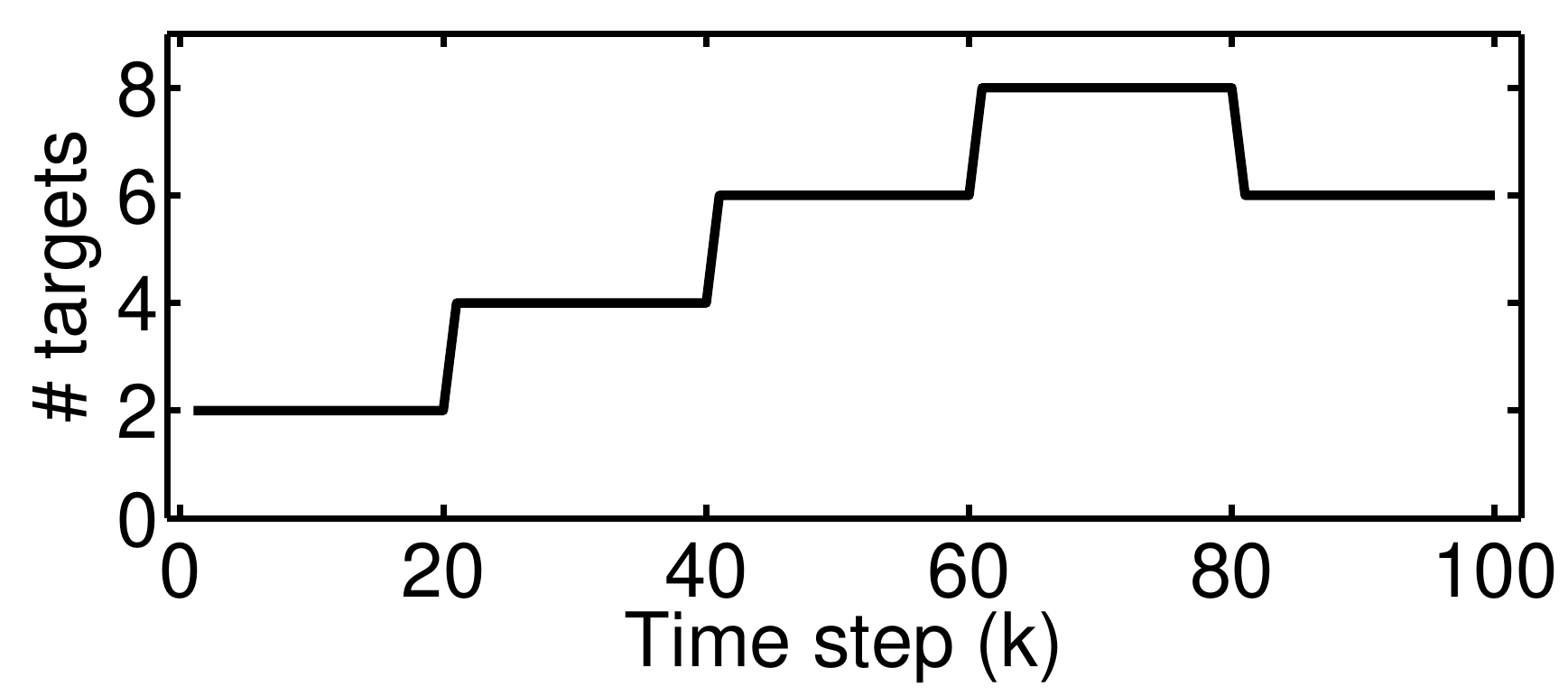}
			\label{fig:card_8t}}
		
		\caption{(a): Evolution of target tracks. The blue cross indicates
			origin of target trajectory. (b): Number of targets as function of time.}
\end{figure}

The single target state is a four dimensional vector $\x = [x, y, v_x, v_y]$
consisting of its position coordinates $x$ and $y$ and its velocities 
$v_x$ and $v_y$ along the $x$-axis and $y$-axis respectively. The target 
state evolves according to the discretized version of the continuous time 
nearly constant velocity model~\cite{barshalom2004} given by
\begin{align}
\x_{k+1,i} &= \left[ \begin{array}{cccc} 1&0&T&0\\ 0&1&0&T\\ 0&0&1&0\\ 0&0&0&1 \end{array} \right] \x_{k,i} + \eta_{k+1,i} \\
\eta_{k+1,i} & \sim	\N({\bf 0},\Sigma_{\eta}) \,, \; 
	\Sigma_{\eta} = \left[ \begin{array}{cccc} \frac{T^3}{3}&0&\frac{T^2}{2}&0 \\ 
	0&\frac{T^3}{3}&0&\frac{T^2}{2} \\ \frac{T^2}{2}&0&T&0\\ 0&\frac{T^2}{2}&0&T \end{array} \right] \sigma_{\eta}^{2}
	\nonumber
\end{align}
where $T$ is the sampling period and $\sigma_{\eta}^{2}$ is the intensity
of the process noise. We simulate $100$ time steps with a sampling 
period of $T = 1s$ and process noise intensity of $\sigma_{\eta} = 0.25m$.
Figure~\ref{fig:tracks_8t} shows the target tracks used in the simulations 
and Figure~\ref{fig:card_8t} shows the variation of the number of targets 
over time. All the targets originate from one of the following four locations 
$(\pm 400m, \pm 400m)$ and targets are restricted to the $2000m \times 2000m$ 
square region centered at the origin.
Targets  1 \& 2 are present in the time range $k \in \ll 1 , 100 \rr$; targets 
3 \& 4 for $k \in \ll 21 , 100 \rr$; targets 5 \& 6 for $k \in \ll 41 , 100 \rr$;
and targets 7 \& 8 for $k \in \ll 61 , 80 \rr$.

\subsection{Measurement model}

Measurements are collected independently by six sensors. When a sensor 
detects a target, the corresponding measurement consists of the position
coordinates of the target corrupted by additive Gaussian noise. Thus if 
a target located at $(x,y)$ is detected by a sensor, the measurement 
gathered by the sensor is given by
\begin{align}
\z = \left[\begin{array}{c} x \\ y \end{array}\right] + 
	\left[\begin{array}{c} w_x \\ w_y \end{array}\right]
\end{align}
where $w_x$ and $w_y$ are independent zero-mean Gaussian noise terms with 
standard deviation $\sigma_{w_x}$ and $\sigma_{w_y}$ respectively. In our 
simulations we use $\sigma_{w_x} = \sigma_{w_y} = 10m$. The probability 
of detection of each sensor is constant throughout the monitoring region. 
Five of the sensors have a fixed probability of detection of $0.5$. The 
probability of detection of the sixth sensor is variable and is changed 
from $0.2$ to $1$ in increments of $0.1$. The clutter measurements made 
by each of the sensors is Poisson with uniform spatial density and mean 
clutter rate $\lambda = 10$.

\subsection{Filter implementation details and error metric}
\label{sec:filter_details}

All the filters model the survival probability at all times and at all 
locations as constant with $p_{sv} = 0.99$. The target birth intensity 
is modelled as a Gaussian mixture with four components centered at 
$(\pm 400, \pm 400, 0, 0)$, each with covariance matrix 
diag$([100,100,25,25])$ and weight $0.1$. The target birth cardinality 
distribution is assumed Poisson with mean $0.4$. We consider two cases 
of sensor ordering where the sensor with variable probability of detection 
is either processed first (Case 1) or last (Case 2).

For the different multisensor filters the PHD function is represented 
by a mixture of Gaussian densities whereas the cardinality distribution 
is represented by a vector of finite length which sums to one. This 
Gaussian mixture model approximation was first used in~\cite{vo2006} 
and~\cite{vo2007} for multitarget tracking using single sensor PHD 
and CPHD filters respectively.
We perform pruning of Gaussian components with low weights and merging 
of Gaussian components in close vicinity~\cite{vo2006} for computational 
tractability. For the iterated-corrector filters pruning and merging is 
done after processing each sensor since many components have negligible 
weight and propagating them has no significant impact on tracking 
accuracy. For the general multisensor PHD and CPHD filters pruning and 
merging is performed at the end of the update step since intermediate 
Gaussian components are not accessible.
The general multisensor PHD and CPHD filters are implemented using the 
two-step greedy approach described in Section~\ref{sec:apprx_implementation}.
In our simulations the maximum number of measurement subsets per Gaussian 
component is set to $W_{\textrm{max}} = 6$ and the maximum number of partitions
of measurement subsets is set as $P_{\textrm{max}} = 6$. For CPHD filters, 
the cardinality distribution is assumed to be zero for $n > 20$.

For the PHD filters, we estimate the number of targets by rounding the 
sum of weights of the Gaussian components to the nearest integer. For 
the CPHD filters, we estimate the number of targets as the peak of the 
posterior cardinality distribution. For all the filters, the target 
state estimates are the centres of the Gaussian components with highest 
weights in the posterior PHD. To reduce the computational overhead, 
after each time step we restrict the number of Gaussian components to 
a maximum of four times the estimated number of targets. When the 
estimated number of targets is zero we retain a maximum of four Gaussian
components.

The tracking performance of the different filters are compared using the 
\emph{optimal sub-pattern assignment} (OSPA) error metric~\cite{schuhmacher2008}. 
For the OSPA metric, we set the cardinality penalty factor $c = 100$ and
power $p = 1$. The OSPA error metric accounts for error in estimation of 
target states as well as the error in estimation of number of targets. 
Given the two sets of estimated multitarget state and the true multitarget 
state, it finds the best permutation of the larger set which minimizes its 
distance from the smaller set and assigns a fixed penalty $c$ for each 
cardinality error. We use the Euclidean distance metric and consider only
the target positions while computing the OSPA error.

\subsection{Results}

\begin{figure}[h!]
	\centering
		\subfigure[Solid lines: Case 1; Dashed lines: Case 2]{
			\includegraphics[width=0.45\textwidth]{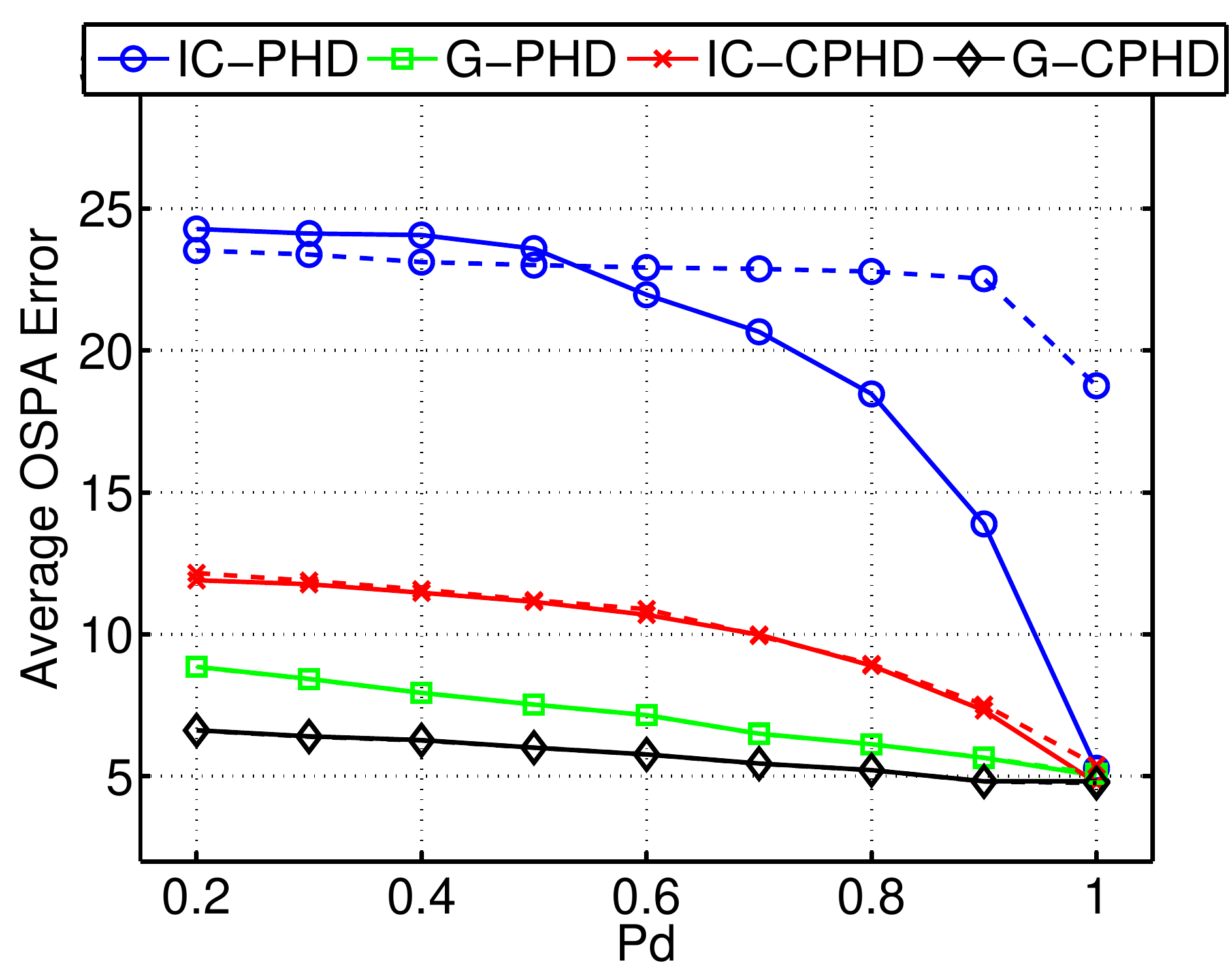}
			\label{fig:plots_ospa_pd}}

		\subfigure[Solid lines: Case 1; Dashed lines: Case 2]{
			\includegraphics[width=0.45\textwidth]{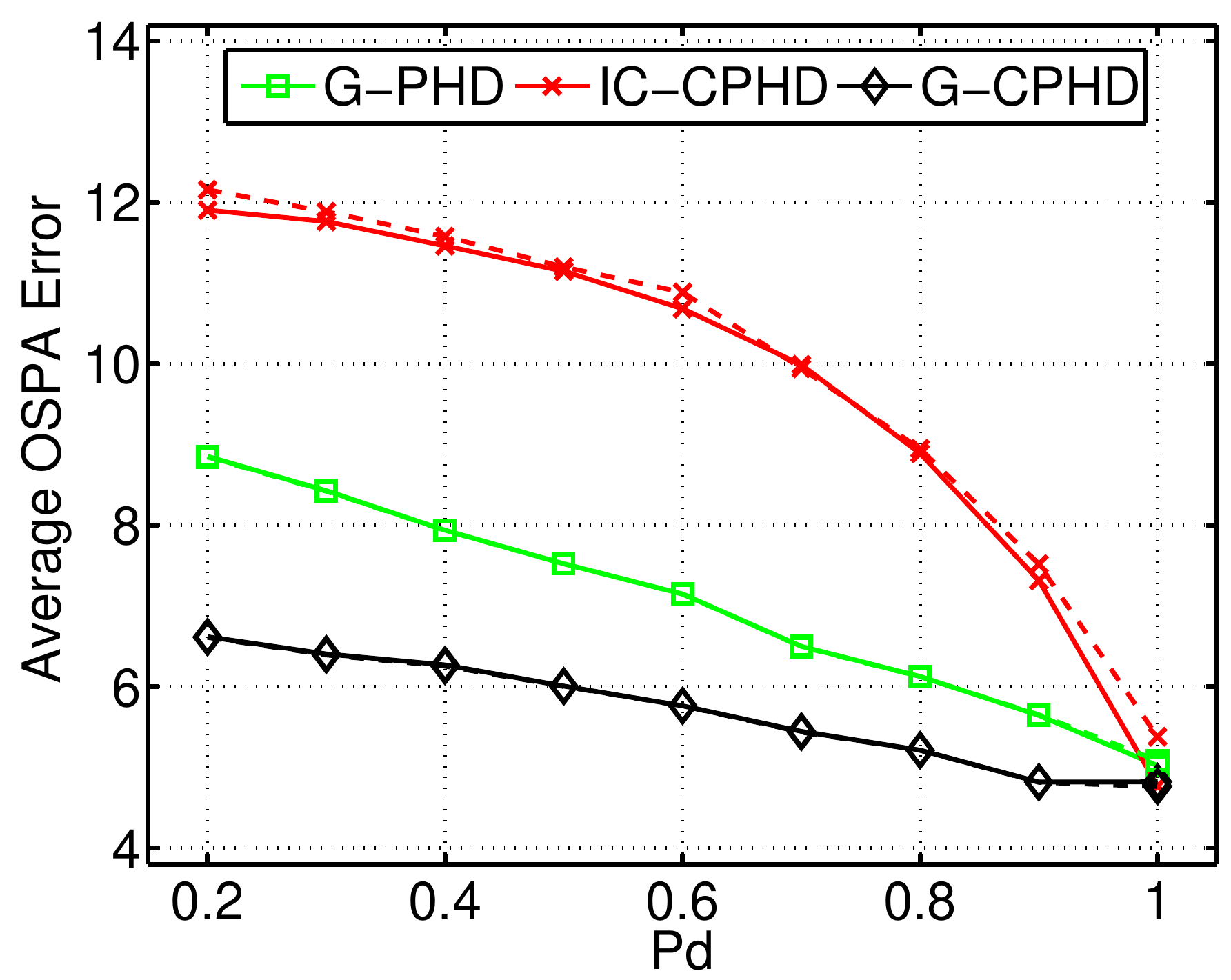}
			\label{fig:plots_ospa_pd_zoom}}
		
		\subfigure[Box and whisker plot]{
			\includegraphics[width=0.43\textwidth]{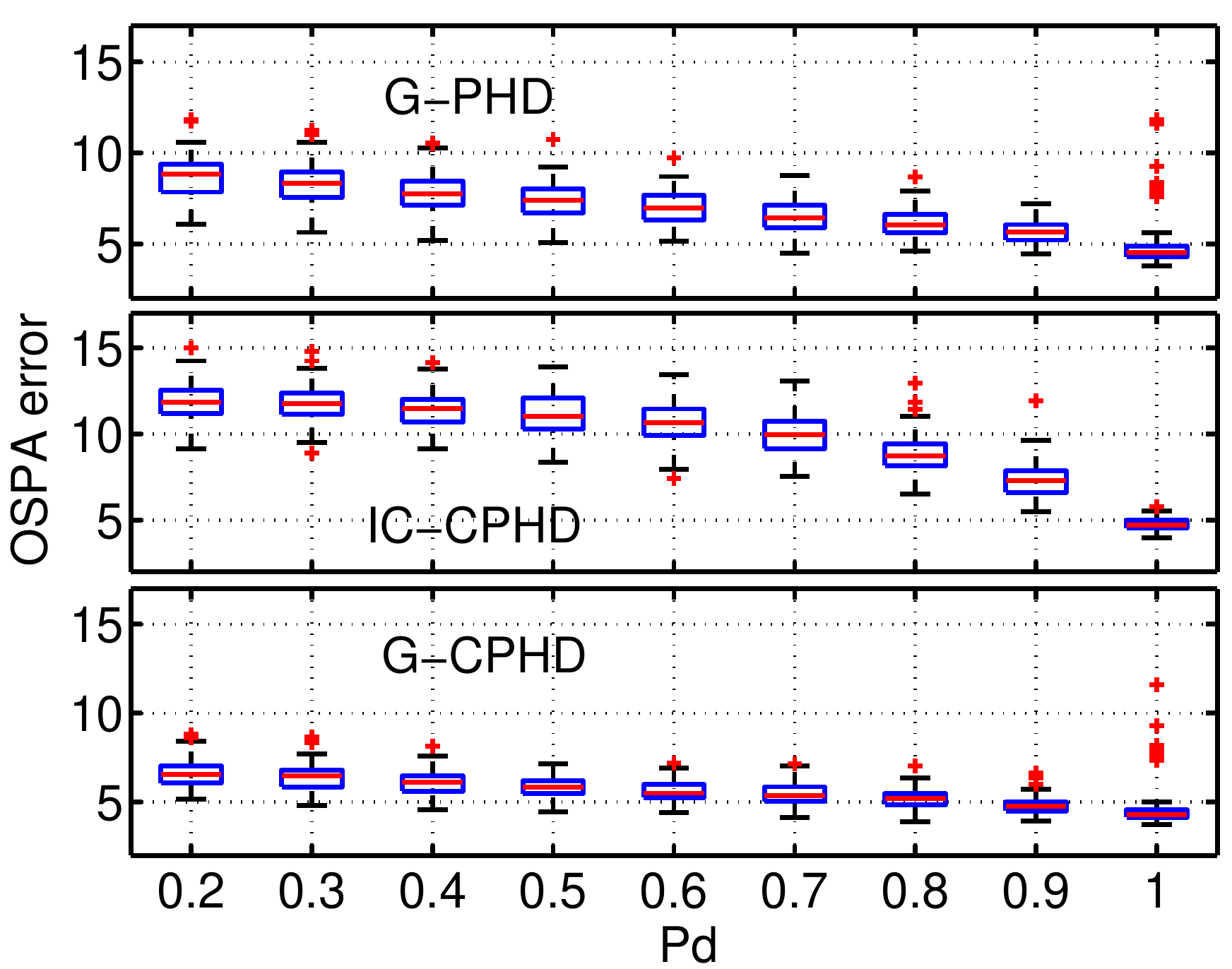}
			\label{fig:plot_box_1}}
			
		\caption{(a): Average OSPA error versus the probability of detection 
				$p_d$ of the variable sensor. The solid and dashed lines correspond 
				to Case 1 and Case 2, respectively.
				(b): A zoomed-in version of the figure in (a) focusing on the IC-CPHD, 
				G-PHD and G-CPHD filters.
				(c): Box and whisker plot of the OSPA error as a function of $p_d$. 
				Boxes indicate 25-75 interquartile range; whiskers extend 1.5 times 
				the range and `+' symbols indicate outliers lying beyond the whiskers.}
\end{figure}

The target tracks shown in Figure~\ref{fig:tracks_8t} are used in all the 
simulations. The generated observation sequence is changed by providing a 
different initialization seed to the random number generator. We generate 
100 different observation sequences and report the average OSPA error 
obtained by running each multisensor filter over these 100 observation 
sequences. The probability of detection of the sensor with variable 
probability of detection is gradually increased from $0.2$ to $1$. 
Figure~\ref{fig:plots_ospa_pd} shows the average OSPA error as the 
probability of detection is changed for the two cases, Case 1 and Case 2.
The IC-PHD filter performs significantly worse than all the other filters.
For the IC-PHD filter Case 1, the accuracy improves relative to Case 2 as 
the probability of detection is increased since the sensor with more
reliable information is processed towards the end.

Figure~\ref{fig:plots_ospa_pd_zoom} shows a portion of 
Figure~\ref{fig:plots_ospa_pd} enlarged for clarity. We observe that for the 
G-PHD, IC-CPHD and G-CPHD filters there is very little difference between 
performance for Case 1 and Case 2. Thus the IC-CPHD filter performance does 
not depend significantly on the order in which sensors are processed. For 
the G-PHD and G-CPHD filters the order in which sensors are processed to 
greedily construct measurement subsets has little impact on the final filter 
performance. The G-CPHD filter is able to outperform both the G-PHD and the 
IC-CPHD filters and has the lowest average OSPA error. 
A box and whisker plot comparison of the  G-PHD, IC-CPHD and G-CPHD 
filters is shown in Figure~\ref{fig:plot_box_1}. The median OSPA error and 
the $25-75$ percentiles are shown for different values of $p_d$ for the 
sensor with variable probability of detection.

We now examine the effect of the parameters $W_{\text{max}}$ and 
$P_{\text{max}}$, i.e., the maximum number of measurement subsets and the 
maximum number of partitions. $W_{\text{max}}$ is varied 
in the range $\{1,2,4,6,8,10\}$ and $P_{\text{max}}$ is varied in the range
$\{1,2,4,6,8,10\}$. For this simulation we fix the probability of detection 
of all the six sensors to be $0.5$. All other parameters of the simulation 
are the same as before. We do tracking using the same tracks as before and 
over 100 different observation sequences for each pair of 
$(W_{\text{max}}, P_{\text{max}})$.

\begin{figure}[h!]
	\centering
		\subfigure[Solid lines: G-CPHD; Dashed lines: G-PHD]{
			\includegraphics[width=0.43\textwidth]{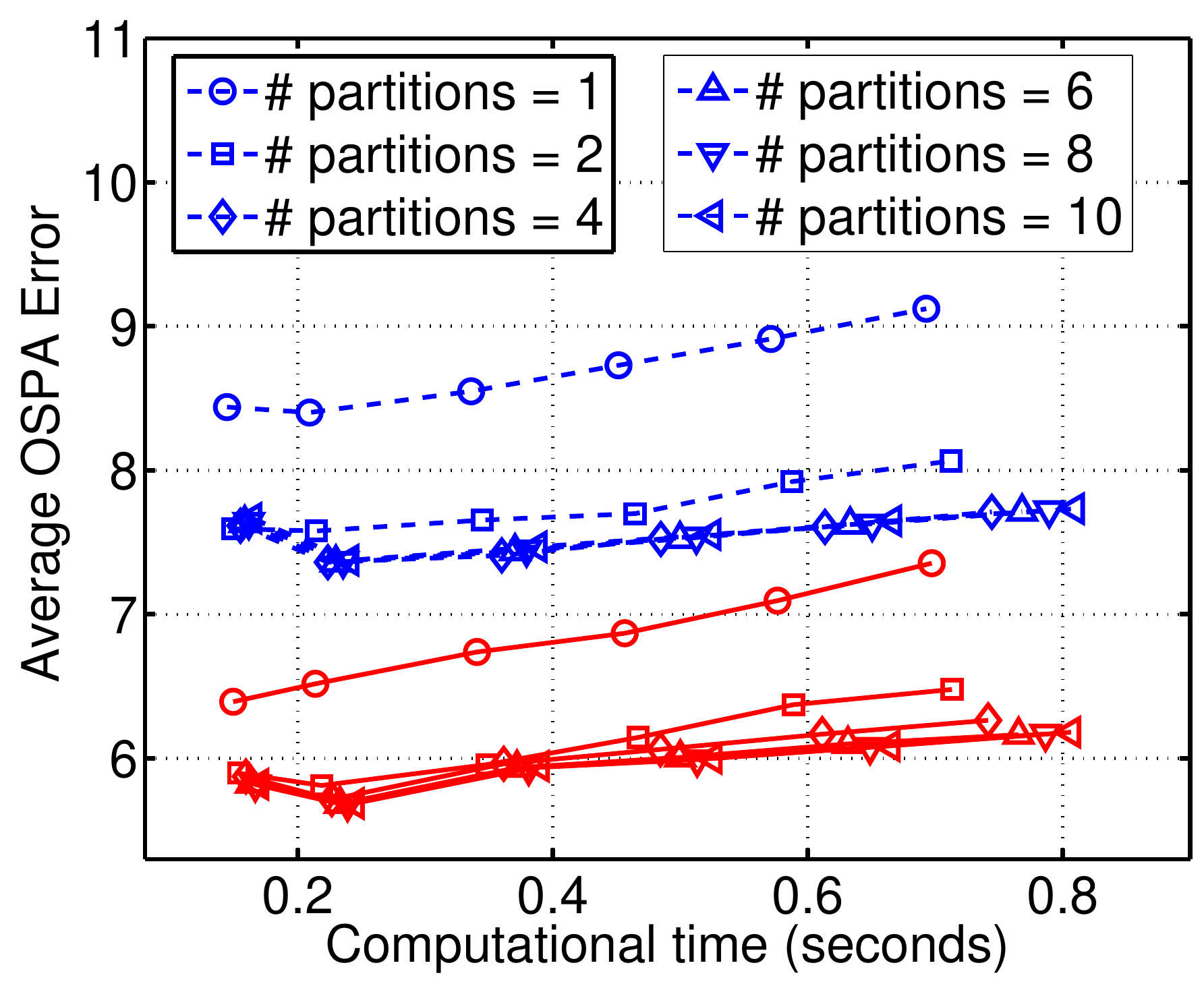}
			\label{fig:plot_error_time}}
			
		\subfigure[Solid lines: G-CPHD; Dashed lines: G-PHD]{
			\includegraphics[width=0.44\textwidth]{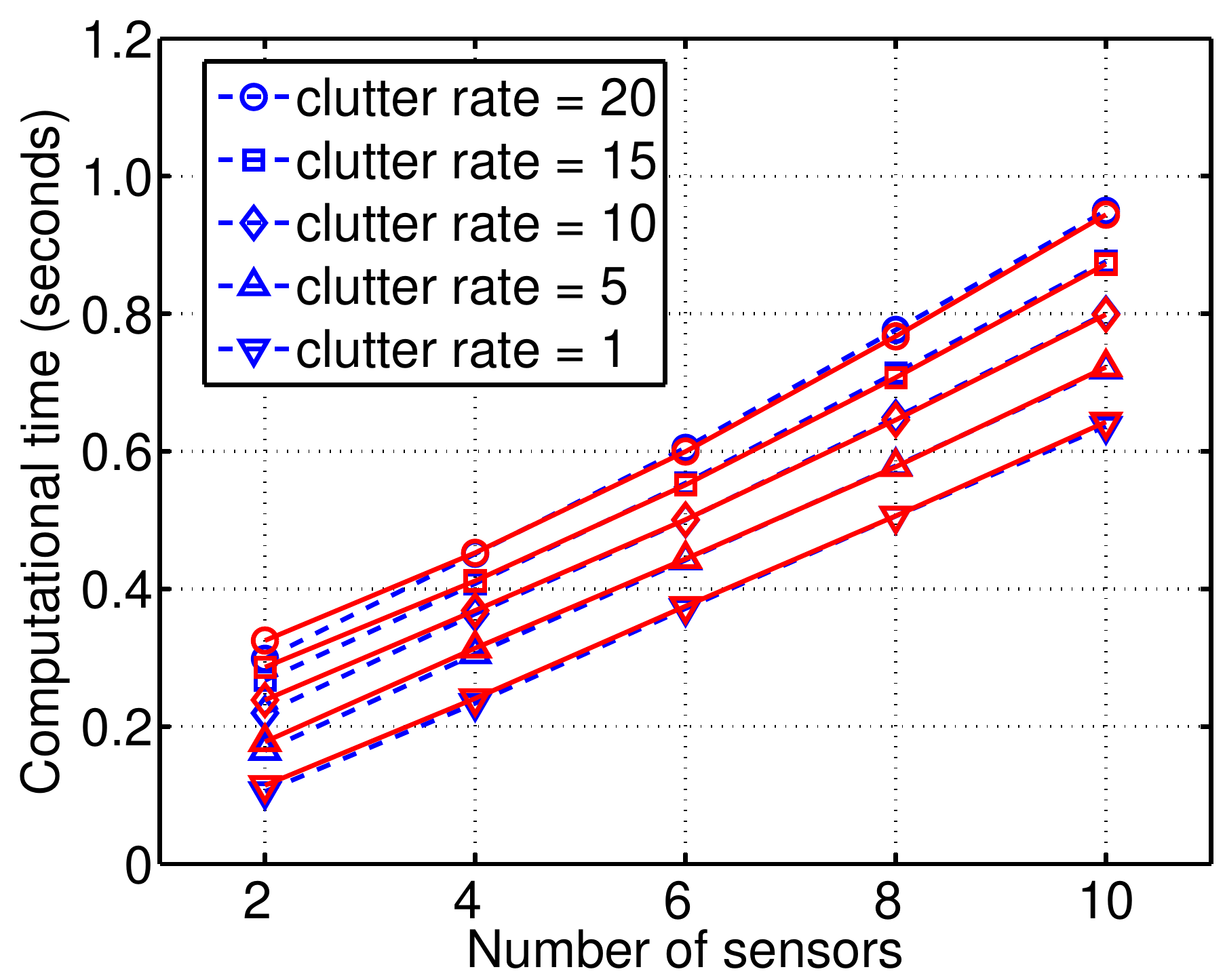}
			\label{fig:plots_sensor_1}}

		\subfigure[Solid lines: G-CPHD; Dashed lines: G-PHD]{
			\includegraphics[width=0.44\textwidth]{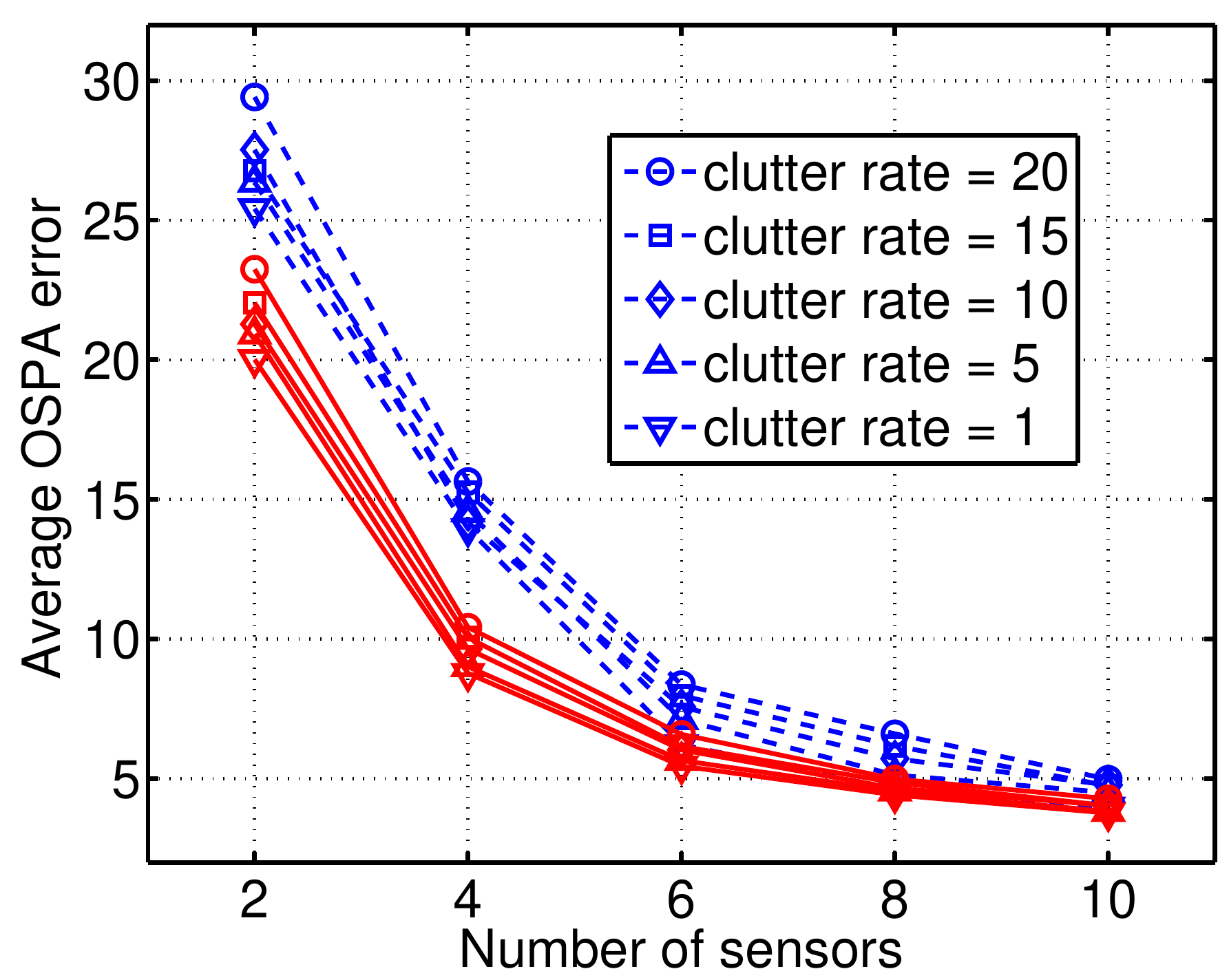}
			\label{fig:plots_sensor_2}}
		
		\caption{(a): Average OSPA error versus computational
                  time per time-step obtained by changing 
				$W_{\text{max}}$ in the range $\{1,2,4,6,8,10\}$ and 
				$P_{\text{max}}$ in the range $\{1,2,4,6,8,10\}$. Blue dashed 
				curves correspond to G-PHD filter and red solid curves 
				correspond to G-CPHD filter.
				(b): Computation time required as a function of number of sensors. 
				(c): Average OSPA error as a function of number of sensors.}
\end{figure}

Figure~\ref{fig:plot_error_time} plots the effect of changing
$W_{\text{max}}$ and $P_{\text{max}}$ on the average OSPA error and
the average computational time required per time-step. Simulations
were performed using algorithms implemented in Matlab on computers
with two Xeon 4-core 2.5GHz processors and 14GB RAM. Each curve is obtained by fixing
$P_{\text{max}}$ and changing $W_{\text{max}}$. Dashed curves correspond
to G-PHD filter and solid curves correspond to G-CPHD filters. For a given 
pair of $(W_{\text{max}}, P_{\text{max}})$ values both the filters require 
almost the same computational time but the G-CPHD filter has a lower
average OSPA error compared to the G-PHD filter.
We observe that for each curve as $W_{\text{max}}$ increases the average
OSPA error reaches a minimum quickly (around $W_{\text{max}} = 2$) and
then starts rising. This is because as $W_{\text{max}}$ is increased the 
non-ideal measurement subsets also get involved in the construction of 
partitions leading to noise terms in the update. The computational
time required grows approximately linearly with increase in $W_{\text{max}}$.
As $P_{\text{max}}$ is varied the average OSPA error saturates at around 
$P_{\text{max}} = 4$ and increasing it beyond 4 has very little impact. 
Increasing $P_{\text{max}}$ does not significantly raise the computational 
time requirements of the approximate G-PHD and G-CPHD filter implementations.

We perform another set of simulations to study the effect of the number of 
sensors and clutter rate of the sensors on filter performance. In this 
simulation the number of sensors $s$ is varied in the range $\{2,4,6,8,10\}$ 
and the clutter rate $\lambda$ of the Poisson clutter process is varied in 
the range $\{1,5,10,15,20\}$ and is same for all the sensors. We fix the 
probability of detection of all the sensors to be $0.5$. The approximate 
greedy algorithm parameters are set to $W_{\text{max}} = 6$ and 
$P_{\text{max}} = 6$. All other parameters of the simulation are unchanged. 
Tracking is performed using the same target tracks as before and 100 different 
observation sequences are generated for each pair of $(s,\lambda)$.

Figures~\ref{fig:plots_sensor_1} and~\ref{fig:plots_sensor_2} plot the 
average computational time and the average OSPA error as the number of 
sensors is changed for different clutter rate values. Each curve is 
obtained by fixing the clutter rate and changing the number of sensors. 
Dashed curves correspond to the G-PHD filter and solid curves correspond 
to the G-CPHD filters. From Figure~\ref{fig:plots_sensor_1} we observe 
that for approximate greedy implementations of the G-PHD and G-CPHD 
filters the computational requirements grow linearly with the number of 
sensors. As the number of sensors is increased the average OSPA error 
reduces as seen from Figure~\ref{fig:plots_sensor_2}. The G-CPHD filter 
requires relatively fewer sensors to achieve the same accuracy as that 
of the G-PHD filter.

\subsection{Extension to non-linear measurement model}

In this section we extend the Gaussian mixture based filter implementation 
discussed in Section~\ref{sec:apprx_implementation} to include non-linear
measurement models using the unscented Kalman filter~\cite{julier1997} approach.
The unscented extensions to non-linear models when a single sensor is present 
are discussed in~\cite{vo2006} and~\cite{vo2007} for the PHD and CPHD filters 
respectively. We implement the unscented versions of the general multisensor 
PHD and CPHD filters by repeatedly applying the equations provided 
in~\cite{vo2006,vo2007}. Specifically, the equations are recursively applied 
for each $\z \in W$ to evaluate the score function $\beta^{(i)}(W)$ while 
constructing the measurement subsets.

\begin{figure}[h!]
	\centering
		\subfigure[Solid lines: G-CPHD; Dashed lines: G-PHD]{
			\includegraphics[width=0.45\textwidth]{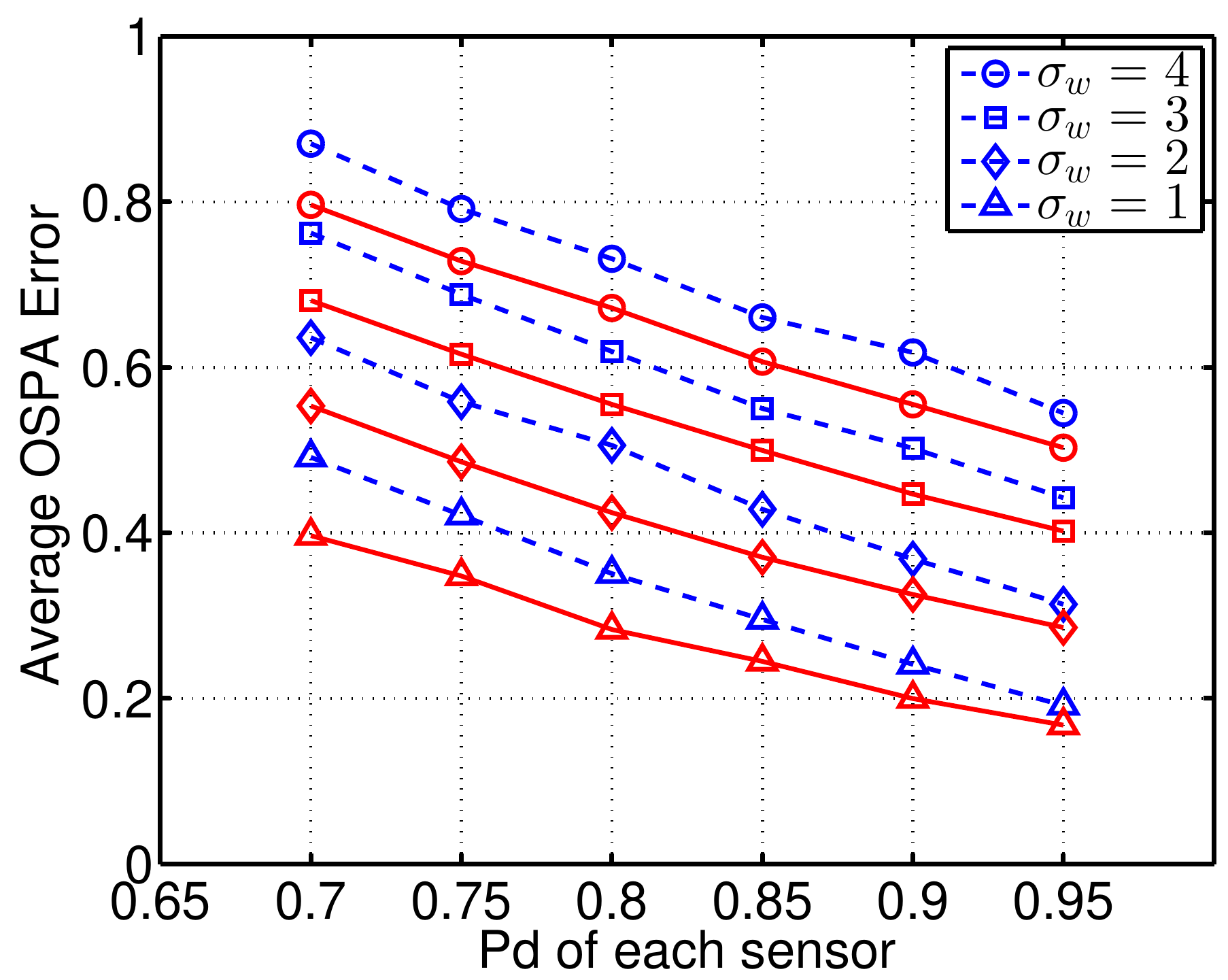}
			\label{fig:avg_ospa_drdc}}
			
		\subfigure[Solid lines: True tracks; Dashed lines: Estimated tracks]{
			\includegraphics[width=0.43\textwidth]{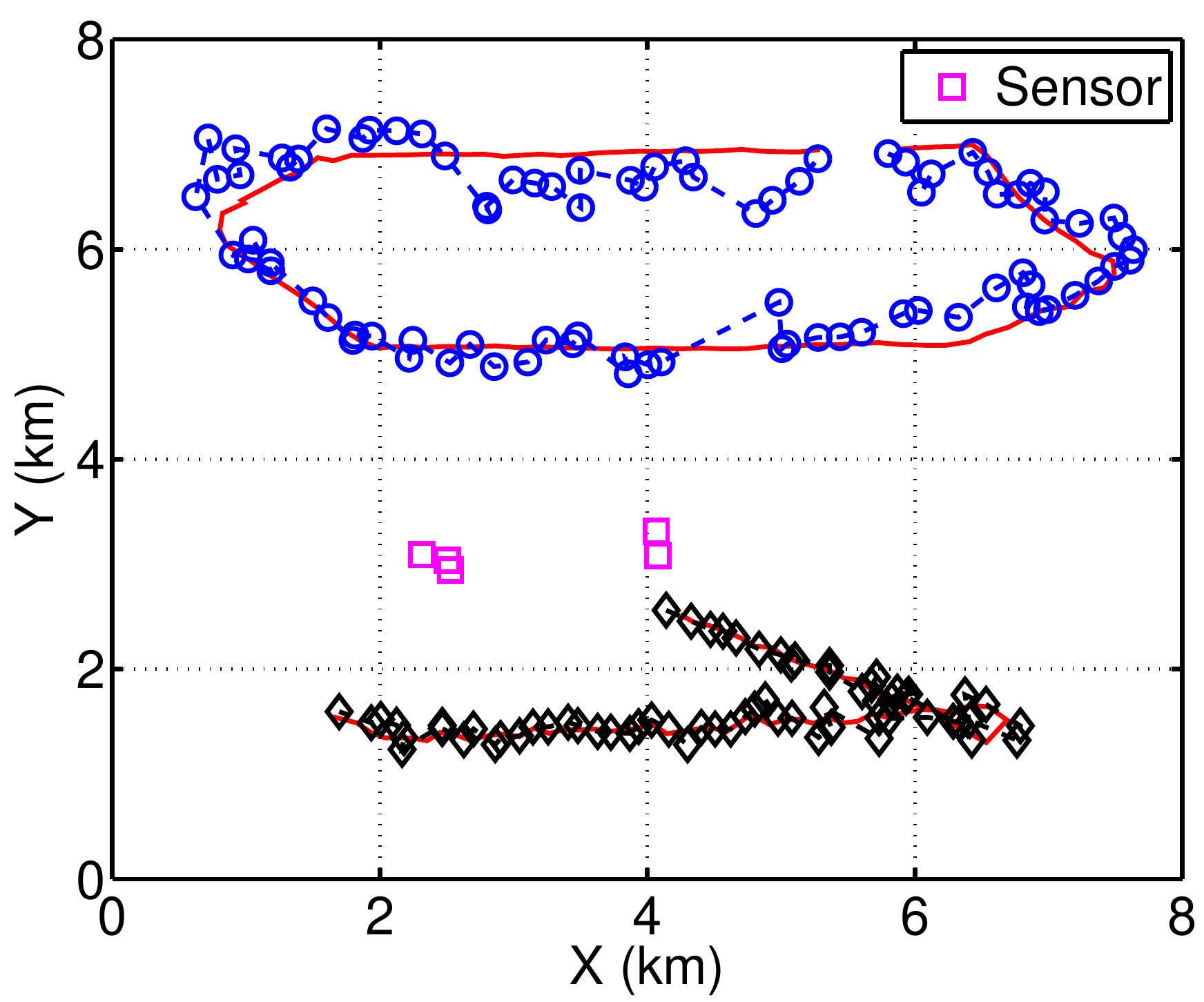}
			\label{fig:tracks_drdc}}
		
		\caption{(a): Average OSPA error Vs probability of detection of 
				individual sensors. Different plots obtained by changing 
				$\sigma_{w}$ in the range $\{1,2,3,4\}$. Blue dashed curves 
				correspond to G-PHD filter and red solid curves correspond to 
				G-CPHD filter.
				(b): True target tracks (solid red) and estimated target tracks 
				(dashed with markers) obtained using the G-CPHD filter when 
				$\sigma_{w} = 2$ and $p_d = 0.9$.}
\end{figure}

As an example, we consider the setup described in~\cite{yu2013} based on 
at-sea experiments. Two targets are present within the monitoring region 
and portions of their tracks are shown in Figure~\ref{fig:tracks_drdc}. 
The target state $\x = [x ,y]$ consists of its coordinates in the $x-y$ 
plane and the filters model the motion of individual targets using a 
random walk model given by $\x_{k+1,i} = \x_{k,i} + \eta_{k+1,i}$ where 
the process noise $\eta_{k+1,i}$ is zero-mean Gaussian with covariance 
matrix $\Sigma_{\eta} = \sigma_{\eta}^{2} \, \text{diag}(1,1)$. In our 
simulations we set $\sigma_{\eta} = 0.24 \, \text{km}$. Although we 
consider linear target dynamics in this paper, the unscented approach 
can be easily extended to include non-linear target dynamics as well.

The targets are monitored using acoustic sensors which collect bearings (angle) 
measurements. If sensor $j$ is present at location $[x^{j}, y^{j}]$ and a target 
detected by the sensor has coordinates $ [x, y]$ then the measurement made by 
this sensor is given by
\begin{align}
z &= \text{arctan} \left(\frac{y - y^{j}}{x - x^{j}}\right) + w
\end{align}
where the measurement noise $w$ is zero mean Gaussian with standard deviation 
$\sigma_{w}$ and `arctan' denotes the four-quadrant inverse tangent function. 
The sensor locations are assumed to be known. The measurements 
$z$ are in the range $[0,360)$ degrees. Along with the target related 
measurements the sensors also record clutter measurements not associated
with any target. Five sensors (which slowly drift over time) gather measurements 
and their approximate locations are indicated in Figure~\ref{fig:tracks_drdc}. 
To demonstrate the feasibility of the proposed algorithms for non-linear 
measurement models, in this paper we consider sensor deployments and target 
trajectories based on at-sea experiments. The sensor measurements themselves 
are simulated to avoid the issue of measurement model mismatch.
All sensors are assumed to have the same $\sigma_{w}$ and we vary $\sigma_{w}$ 
in the range $\{1,2,3,4\}$ (degrees) in our simulations. The probability of 
detection of each sensor is uniform throughout the monitoring region and is 
the same for all the sensors. The probability of detection of the sensors is 
changed from $0.7$ to $0.95$ in increments of $0.05$. The clutter measurements 
made by each of the sensors is a Poisson random finite set with uniform 
density in $[0,360)$ and mean clutter rate $\lambda = 5$.

The general multisensor PHD and the general multisensor CPHD filters are 
used to perform tracking in this setup. Most of the implementation details 
are the same as discussed in Section~\ref{sec:filter_details}. The target birth 
intensity is modeled as a two component Gaussian mixture with components 
centered at true location of the targets at time $k = 1$ and each having 
covariance matrix $\text{diag}([0.65,0.79])$ and weight $0.1$. The target 
birth cardinality distribution is assumed to be Poisson with mean $0.2$. 
We set $W_{\textrm{max}} = 6$ and $P_{\textrm{max}} = 6$. While calculating 
the OSPA error we use the cardinality penalty factor of $c = 2$ and power 
$p = 1$. To demonstrate the feasibility of the proposed algorithm, 
in our current implementation a-priori information about initial target 
locations is assumed to be known. In a more practical scenario where this 
information is unavailable the Gaussian components (for birth) can be 
initialized based on sensor measurements.

The average OSPA error obtained by running the algorithms over 100 different
observation sequences are shown in Figure~\ref{fig:avg_ospa_drdc}. Each curve 
is obtained by varying the probability of detection of the sensors from $0.7$
to $0.95$. As the probability of detection increases there is gradual decrease
in the average OSPA error. Different curves correspond to different values of 
measurement noise standard deviation $\sigma_{w}$. As $\sigma_{w}$ is increased 
the average OSPA error increases as expected. For each $\sigma_{w}$ the G-CPHD 
filter performs better than the G-PHD filter. Estimated target locations obtained 
by the general multisensor CPHD filter are shown in Figure~\ref{fig:tracks_drdc}
when $\sigma_{w} = 2$ and $p_d = 0.9$. The tracks are obtained by joining the 
closest estimates across time.

\section{Conclusions}
\label{sec:conclusion}

In this paper we address the problem of multitarget tracking using 
multiple sensors. Many of the existing approaches do not make complete 
use of the multisensor information or are computationally infeasible.
The contribution of this work is twofold. As our first contribution
we derive update equations for the general multisensor CPHD filter. 
These update equations, similar to the general multisensor PHD filter 
update equations, are combinatorial in nature and hence computationally 
intractable. Our second contribution is in developing an approximate 
greedy implementation of the general multisensor CPHD and PHD filters 
based on a Gaussian mixture model. The algorithm avoids any combinatorial
calculations without sacrificing tracking accuracy. The algorithm is 
also scalable since the computational requirements grow linearly in the
number of sensors as observed from the simulations.

\section*{Acknowledgements}
We thank Florian Meyer for pointing out an issue in an earlier version of 
this manuscript. We also thank the anonymous reviewers for comments 
which helped improve the presentation of this work.

\appendices

\section{Recursive expression for partitions}
\label{app:recursive}

Let $\S^{(\l)}$ be the collection of all permissible partitions of the set 
$Z^{1:\l}_{k+1}$ ($1 \leq \l < s$) where partitions are as defined in the
equations~\eqref{eq:partition_1}-\eqref{eq:partition_4}. Since the $V$ 
component of a partition is unique given the $W$ components, we do not 
explicitly specify the $V$ component in the recursive expression. 
Let $P \in \S^{(\l)}$ be any partition of $Z^{1:\l}_{k+1}$ which is given 
as $P = \{W_1,W_2,\dots,,W_{|P|-1}, V \}$. Let 
$Z^{\l+1}_{k+1} = \{ \z^{\l+1}_{1}, \z^{\l+1}_{2}, \dots, \z^{\l+1}_{m_{\l+1}}\}$.
Then we can express $\S^{(\l+1)}$ using $\S^{(\l)}$ and $Z^{\l+1}$ as given by
the following relation 
\begin{align}
& \S^{(\l+1)} = \nonumber \\
& \bigcup_{P \in \S^{(\l)}} \bigcup_{n_{1}=0}^{m_{\l+1}} \bigcup_{n_{2}=0}^{\textrm{min}(m_{\l+1},|P|-1)}
	\!\!\!\!\!\! \bigcup_{\substack{I_{1} \subseteq \ll 1 , m_{\l+1} \rr \\ |I_{1}| = n_{1}}}
	\bigcup_{\substack{I_{2} \subseteq \ll 1 , m_{\l+1} \rr \\ |I_{2}| = n_{2} \\ I_{1} \cap I_{2} = \emptyset}}
	\bigcup_{\substack{J \subseteq \ll 1 , |P|-1 \rr \\ |J| = |I_{2}|}}  \nonumber \\
& \quad \bigcup_{B \in \mathcal{B}(I_{2},J)}
	\left\{ \{ \z^{\l+1}_{i_{1}}\}_{i_{1} \in I_{1}} 	\cup 
	\{W_{B(i_{2})}, \z^{\l+1}_{i_{2}}\}_{i_{2} \in I_{2}} \cup  
	\{ W_{j}\}_{j \notin J} \right\}
\end{align}
where $\mathcal{B}(I_{2},J)$ is the collection of all possible 
	matchings~\footnote{$\mathcal{B}(I,J)$ is the collection of all
	possible one-to-one mappings from set $I$ to set $J$.}
from set $I_{2}$ to set $J$.
The above relation mathematically expresses the fact that for each partition 
$P \in \S^{(\l)}$ and given $Z^{\l+1}_{k+1}$, a new partition belonging to 
$\S^{(\l+1)}$ can be constructed by adding some new singleton measurement 
subsets from $Z^{\l+1}_{k+1}$ (i.e. $\{ \z^{\l+1}_{i_{1}}\}_{i_{1} \in I_{1}}$), 
extending some existing subsets in $P$ by appending them with measurements 
from $Z^{\l+1}_{k+1}$ (i.e. $\{W_{B(i_{2})} \cup \z^{\l+1}_{i_{2}}\}_{i_{2} \in I_{2}}$)
and retaining some existing measurement subsets (i.e. $\{ W_{j}\}_{j \notin J}$).
By this definition we have $\S = \S^{(s)}$.
As a special case for $\l = 1$,
\begin{align}
\S^{(1)} &= \bigcup_{n=0}^{m_{1}} \bigcup_{\substack{I \subseteq \ll 1 , m_{1} \rr \\ |I| = n}}
	\left\{ \{ \z^{1}_{i}\}_{i \in I} \right\}. \label{eq:union_r1}
\end{align}

\section{Functional derivatives}
\label{app:functional_derivatives}

We now review the notion of functional derivatives which play an 
important role in the derivation of filter update equations. A brief 
background is provided that is necessary for the derivations in this 
paper; for additional details see~\cite{mahler2007}, 
\cite[Ch.~11]{mahler2007B}, \cite{mahler2014B}. Let $\F$ denote the set of mappings from 
$\Y$ to $\R$ and let $A$ be a functional mapping elements of $\F$ to $\R$.
Let $u(\y)$ and $g(\y)$ be functions in $\F$. For the functional 
$A[u]$, its G\^{a}teaux derivative along the direction of the function 
$g(\y)$ is defined as~\cite{mahler2007B}
\begin{align}
\frac{\partial A}{\partial g}[u] \, \eqdef \, \lim_{\epsilon \rightarrow 0}
	\frac{A[u + \epsilon \cdot g] - A[u]}{\epsilon}.
\end{align}
In this paper we are interested in the G\^{a}teaux derivatives when 
the function $g(\y)$ is the Dirac delta function $\delta_{\y_{1}}(\y)$ 
localized at $\y_{1}$ and the corresponding G\^{a}teaux derivatives are 
called functional derivatives \cite{mahler2007B,mahler2007}. 
In this case, the functional derivative is commonly written,
\begin{align}
\frac{\partial A}{\partial \delta_{\y_{1}}}[u] \equiv \frac{\delta A}{\delta \y_{1}}[u].
\end{align}
If the functional $A[u]$ is of the form $A[u] = \int \! {u(\y) g(\y) d\y}$ 
then we have $\frac{\delta A}{\delta \y_{1}}[u] = g(\y_{1})$.

We can also define higher order functional derivatives of $A[u]$. For a set 
$Y = \{ \y_{1}, \y_{2}, \cdots, \y_{n} \}$, the $n^{th}$ order derivative
is denoted by
\begin{align}
\frac{\delta^{n} A}{\delta Y}[u] &\eqdef 
	\frac{\delta^{n} A}{\delta \y_{1} \delta \y_{2} \dots \delta \y_{n}}[u].
\end{align}
We call $\frac{\delta^{n} A}{\delta Y}[u]$ the functional derivative of 
$A[u]$ with respect to the set $Y$.

For functionals $A_{1}[u]$, $A_{2}[u]$ and $A_{3}[u]$, the product rule for 
functional derivatives~\cite[Ch.~11]{mahler2007B} gives
\begin{align}
&\frac{\delta}{\delta Y} \{A_{1}[u] \, A_{2}[u] \, A_{3}[u]\} \nonumber \\ 
&= \sum_{Y_{1} \subseteq Y} \sum_{\substack{Y_{2} \subseteq Y \\ Y_{1} \cap Y_{2} = \emptyset}} 
	\frac{\delta A_{1}}{\delta Y_{1}}[u] \frac{\delta A_{2}}{\delta Y_{2}}[u] 
	\frac{\delta A_{3}}{\delta (Y-Y_{1}-Y_{2})}[u] \\
&= \sum_{n_{1}=0}^{|Y|} \sum_{n_{2}=0}^{|Y|} \sum_{\substack{Y_{1} \subseteq Y \\ |Y_{1}| = n_{1}}}
	\sum_{\substack{Y_{2} \subseteq Y \\ |Y_{2}| = n_{2} \\ Y_{1} \cap Y_{2} = \emptyset}}
	\frac{\delta A_{1}}{\delta Y_{1}}[u] \frac{\delta A_{2}}{\delta Y_{2}}[u] 
	\frac{\delta A_{3}}{\delta (Y-Y_{1}-Y_{2})}[u].
	\label{eq:product_rule_3}
\end{align}
As a special case, for the product of two functionals we have
\begin{align}
\frac{\delta}{\delta Y} \{A_{1}[u] \, A_{2}[u]\} &= 
	\sum_{n=0}^{|Y|} \sum_{\substack{Y_{1} \subseteq Y \\ |Y_{1}| = n}}
	\frac{\delta A_{1}}{\delta Y_{1}}[u] \frac{\delta A_{2}}{\delta (Y-Y_{1})}[u].
	\label{eq:product_rule_2}
\end{align}

\section{Probability generating functional}
\label{app:pgfl}

Let $u(\x)$ be a function with the mapping $u:\mathcal{X} \rightarrow [0,1]$,
where $\mathcal{X}$ is the single target state space. For a set 
$X \subseteq \mathcal{X}$, define $u^X \eqdef \prod_{\x \in X}{u(\x)}$. 
Let $\Xi$ be a random finite set with elements in $\mathcal{X}$ and let
$f_{\Xi}(X)$ be its probability density function. The \emph{probability 
generating functional} (PGFL~\cite{mahler2007B}) of the random finite set 
$\Xi$ is defined as the following integral transform 
\begin{align}
G_{\Xi}[u] &\eqdef \int{u^X \; f_{\Xi}(X) \delta X}
\end{align}
where the integration is a set integral~\cite{mahler2007B}.

For a constant $a \in [0,1]$ denote by $u(\x) \equiv a$ the constant function 
$u(\x) = a$, for all $\x \in \mathcal{X}$. Let $A[a]$ denote the value
of the functional $A[u]$ evaluated at $u(\x) \equiv a$. 
Recall that for a random variable its moments are related to the derivatives 
of its PGF. Similarly, the first moment or the PHD function of a random finite
set is related to the functional derivative of its PGFL. For the random finite 
set $\Xi$, its PHD function $D_{\Xi}(\x)$ is related to the functional 
derivative of the PGFL~\cite{mahler2007B} as follows 
\begin{align}
D_{\Xi}(\x) = \frac{\delta G_{\Xi}}{\delta \x}[1]. \label{eq:phd_functional_derivative}
\end{align}

The PGF $\pgf_{\Xi}(t)$ of the cardinality distribution of the random finite 
set $\Xi$ is obtained by substituting the constant function $u(\x) \equiv t$, 
in the PGFL i.e., $\pgf_{\Xi}(t) = G_{\Xi}[t]$. For an IIDC random finite 
set with spatial density function $\zeta(\x)$ we have the relation
$G_{\Xi}[u] = \pgf_{\Xi}(\Dotp{\zeta}{u})$.

\section{Multitarget Bayes filter}
\label{app:multitarget_bayes}

Let $f_{k+1|k}(X|Z^{1:s}_{1:k})$ and $f_{k+1|k+1}(X|Z^{1:s}_{1:k+1})$ be the 
predicted and posterior multitarget state distributions at time $k+1$ and let 
$L_{k+1,j}(Z^{j}_{k+1}|X)$ denote the multitarget likelihood function for the 
$j^{th}$ sensor at time $k+1$. Since the sensor observations are independent
conditional on the multitarget state, the update equation for the multitarget 
Bayes filter~\cite{mahler2007B} is given by
\begin{align}
f_{k+1|k+1}(X|Z^{1:s}_{1:k+1}) &\propto f_{k+1|k}(X|Z^{1:s}_{1:k}) 
	\prod_{j=1}^{s}{L_{k+1,j}(Z^{j}_{k+1}|X)}.
\end{align}
We now define a multivariate functional which is the integral transform of 
the quantity in the right hand side of the above equation. Under the conditions 
of Assumption 1, we can obtain a closed form expression for this multivariate 
functional, which on differentiation gives the PGFL of the posterior 
multitarget state distribution.

Let $g_{j}(\z), j = 1,2,\dots,s$ be functions that map the space
$\mathcal{Z}^{j}$ to $[0,1]$ where $\mathcal{Z}^{j}$ is the space 
of observations of sensor $j$. The intermediate functions $g_{j}(\z)$ 
will be used to define functionals and later set to zero to obtain 
the PGFL of the posterior multitarget distribution.
Let $u(\x)$ be a function mapping the state space $\mathcal{X}$ to $[0,1]$. 
For brevity, denote the vector of functions $[g_1,g_2,\dots,g_s]$ as $g_{1:s}$
and define $g_{j}^{Z^{j}} \eqdef \prod_{\z \in Z^{j}}{g_{j}(\z)}$ where
$Z^{j} \subseteq \mathcal{Z}^{j}$. We define the multivariate functional 
$F[g_1,g_2,\dots,g_s,u]$ as the following integral transform
\begin{align}
&F[g_{1:s},u] \eqdef \int{u^{X} \, \left( \prod_{j=1}^{s}{\mathcal{L}_{k+1,j}[g_{j}|X]} \right) \, 
	f_{k+1|k}(X|Z^{1:s}_{1:k}) \, \delta X} \label{eq:multivariate_functional} \\
&\text{where } \mathcal{L}_{k+1,j}[g_{j}|X] \eqdef \int{g_{j}^{Z^{j}} \, L_{k+1,j}(Z^{j}|X) \delta Z^{j}}.
\end{align}
Later we will relate the PGFL of the posterior multitarget distribution
to the derivatives of the functional $F[g_{1:s},u]$ with respect to the
sensor observations $Z^{1:s}_{k+1}$.
Recall that $c_{j}(\z)$ denotes the clutter spatial distribution and $C_{j}(t)$ 
denotes the PGF of the clutter cardinality distribution for the $j^{th}$ sensor.
Under Assumption 1 it can be shown that~\cite{mahler2003} 
\begin{align}
\mathcal{L}_{k+1,j}[g_{j}|X] &= \int{g_{j}^{Z^{j}} \, L_{k+1,j}(Z^{j}|X) \delta Z^{j}} \\
&= C_{j}(\Dotp{c_{j}}{g_{j}}) \, \phi_{g_{j}}^{X} \,, \\
\textrm{where } \phi_{g_{j}}(\x) &\eqdef 1 - p^{j}_{d}(\x) + p^{j}_{d}(\x) \, p_{g_{j}}(\x) \\
p_{g_{j}}(\x) &\eqdef \int{g_{j}(\z) h_{j}(\z|\x) d\z}.
\end{align}

Let $G_{k+1|k}[u]$ denote the PGFL of the predicted multitarget distribution.
Using the above relations in~\eqref{eq:multivariate_functional} we have
\begin{align}
F[&g_{1:s},u] = \int \!\! {u^{X} \, \left( \prod_{j=1}^{s}{C_{j}(\Dotp{c_{j}}{g_{j}})} \,
	\phi_{g_{j}}^{X} \right) \, f_{k+1|k}(X|Z^{1:s}_{1:k}) \, \delta X}
\end{align}
Since both $u$ and $\phi_{g_{j}}$ are functions defined over the
space $\mathcal{X}$, we can combine the product of $u^{X}$ and 
$\prod_{j=1}^{s}{\phi_{g_{j}}^{X}}$ and write 
$(u \, \prod_{j=1}^{s}{\phi_{g_{j}}})^{X}$. Hence we have
\begin{align}
& F[g_{1:s},u] \nonumber \\
&= \left( \prod_{j=1}^{s}{C_{j}(\Dotp{c_{j}}{g_{j}})} \right) \, 
	\int{\left(u \, \prod_{j=1}^{s}{\phi_{g_{j}}} \right)^{X} \, f_{k+1|k}(X|Z^{1:s}_{1:k}) \, \delta X} \\
&= \left( \prod_{j=1}^{s}{C_{j}(\Dotp{c_{j}}{g_{j}})} \right) \, G[u\prod_{j=1}^{s}{\phi_{g_{j}}}] \\	
&= \left( \prod_{j=1}^{s}{C_{j}(\Dotp{c_{j}}{g_{j}})} \right) \, \pgf(\Dotp{r}{u\prod_{j=1}^{s}{\phi_{g_{j}}}}).
\end{align}
The last two steps result from the definition of the PGFL and 
the assumption that the predicted multitarget distribution 
$f_{k+1|k}(X|Z^{1:s}_{1:k})$ is IIDC.

Let $G_{k+1|k+1}[u]$ be the PGFL of the multitarget density 
$f_{k+1|k+1}(X|Z^{1:s}_{1:k+1})$, and let $D_{k+1|k+1}(\x)$ be the posterior 
PHD function. From~\cite{mahler2009a,delande2010} we have the following relation
\begin{align}
G_{k+1|k+1}[u] &= \frac{\displaystyle \frac{\delta F}{\delta Z^{1:s}_{k+1}}[0,0,\dots,0,u]}
	{\displaystyle \frac{\delta F}{\delta Z^{1:s}_{k+1}}[0,0,\dots,0,1]}.
	\label{eq:G_defn_gen}
\end{align}
Since the PHD is the functional derivative of the PGFL, 
from~\eqref{eq:phd_functional_derivative}
\begin{align}
D_{k+1|k+1}(\x) &= 
	\frac{\displaystyle \frac{\delta F}{\delta \x \, \delta Z^{1:s}_{k+1}}[0,0,\dots,0,1]}
	{\displaystyle \frac{\delta F}{\delta Z^{1:s}_{k+1}}[0,0,\dots,0,1]}.
	\label{eq:phd_defn_gen}
\end{align}
Note that the differentiation $\frac{\delta}{\delta Z^{j}_{k+1}}$ is with 
respect to the function variable $g_{j}$ and the differentiation $\frac{\delta}{\delta \x}$ 
is with respect to the function variable $u$. The general multisensor CPHD 
filter update equation is derived by evaluating the functional derivatives 
of $F[g_{1:s},u]$ in~\eqref{eq:G_defn_gen} and~\eqref{eq:phd_defn_gen}.

We now define a quantity $\Gamma$ and the functionals
$\Psi_{P}[g_{1:s},u]$ and $\varphi_{W}[g_{1:s},u]$. The functional
derivatives of $F[g_{1:s},u]$ can be expressed in terms of
these quantities. Let
\begin{align}
\Gamma &\eqdef \prod_{j=1}^{s}{\left( \prod_{\z \in Z^{j}_{k+1}}{c_{j}(\z)} \right)}, \label{eq:Gamma} \\
\Psi_{P}&[g_{1:s},u] \nonumber \\ 
&\eqdef \left( \prod_{j=1}^{s} {C^{(m_{j}-|P|_{j})}_{j}(\Dotp{c_{j}}{g_{j}})} \right) \, 
	\pgf^{(|P|-1)}(\Dotp{r}{u \prod_{j=1}^{s}{\phi_{g_{j}}}}). \label{eq:Psi}
\end{align}
For $W \in \mathcal{W}$, let
\begin{align}
&\varphi_{W}[g_{1:s},u] \nonumber \\
&\eqdef \frac{\displaystyle \int \!\! r(\x) u(\x) \left( \prod_{(i,l) \in T_{W}}{\!\!\!\! p_{d}^{i}(\x)\,h_{i}(\z^{i}_{l}|\x)}\right) \, 
	\left( \prod_{j:(j,*) \notin T_{W}}{\!\!\!\!\!\! \phi_{g_{j}(\x)}} \right) d\x}
	{\displaystyle \prod_{(i,l) \in T_{W}}{c_{i}(\z^{i}_{l})}} \cdot
	\label{eq:varphi}
\end{align}
With these definitions we can prove, via mathematical induction, 
the following lemma.

\begin{lem} \label{lem:F_derivative}
Under the conditions of Assumption 1, the functional derivative of 
$F[g_{1:s},u]$ with respect to the multisensor observation set 
$Z^{1:s}_{k+1}$ is given by
\begin{align}
\frac{\delta F}{\delta Z^{1:s}_{k+1}}[g_{1:s},u] &= 
	\Gamma \sum_{P \in \S} \Psi_{P}[g_{1:s},u] \prod_{W \in P}{\varphi_{W}[g_{1:s},u]}
\end{align}
where $\Gamma$, $\Psi_{P}[g_{1:s},u]$ and $\varphi_{W}[g_{1:s},u]$ are as 
defined in~\eqref{eq:Gamma}, \eqref{eq:Psi} and~\eqref{eq:varphi}.
\end{lem}
Lemma~\ref{lem:F_derivative} is proved in Appendix~\ref{app:F_derivative}.

\section{Proof of Lemma~\ref{lem:F_derivative}}
\label{app:F_derivative}

\begin{proof}

The derivation is based on the approach used by Mahler~\cite{mahler2009a}
to derive multisensor PHD filter equations for the two sensor case and 
its extension by Delande et al.\ \cite{delande2010} for the general case 
of $s$ sensors.
\newline
\newline
{\bf Mathematical induction}

We prove using mathematical induction on $1 \leq \l \leq s$ the following 
result,
\begin{align}
\frac{\delta F}{\delta Z^{1:\l}_{k+1}}[g_{1:s},u] &= \Gamma^{(\l)} \sum_{P \in \S^{(\l)}} 
	\Psi_{P}^{(\l)}[g_{1:s},u] \prod_{W \in P}{\varphi_{W}[g_{1:s},u]}
	\label{eq:induction_statement}
\end{align}
where,
\begin{align}
\Gamma^{(\l)} &\eqdef \prod_{j=1}^{\l}{\prod_{\z \in Z^{j}_{k+1}}{c_{j}(\z)}}, \\
\Psi_{P}^{(\l)}&[g_{1:s},u] \eqdef 
	\left( \prod_{j=1}^{\l}{C^{(m_{j}-|P|_{j})}_{j}(\Dotp{c_{j}}{g_{j}})} \right) \nonumber \\
& \times \left( \prod_{j=\l+1}^{s} {C_{j}(\Dotp{c_{j}}{g_{j}})} \right)
	\times \pgf^{(|P|-1)}(\Dotp{r}{u \prod_{j=1}^{s}{\phi_{g_{j}}}}),
\end{align}
and for $W \in \mathcal{W}$,
\begin{align}
&\varphi_{W}[g_{1:s},u] \nonumber \\
&\eqdef \frac{\displaystyle \int \!\! r(\x) u(\x) \left( \prod_{(i,l) \in T_{W}}{\!\!\!\! p_{d}^{i}(\x)\,h_{i}(\z^{i}_{l}|\x)}\right) \, 
	\left( \prod_{j:(j,*) \notin T_{W}}{\!\!\!\!\! \phi_{g_{j}(\x)}} \right) d\x}
	{\displaystyle \prod_{(i,l) \in T_{W}}{c_{i}(\z^{i}_{l})}} \cdot
\end{align}
\newline
{\bf Mathematical induction: case $\l = 1$}

We first establish the induction result for the base case, i.e. $\l = 1$.
Ignoring the time index let the observation set gathered by sensor 1 at time 
$k+1$ be $Z^{1}_{k+1} = \{ \z^{1}_{1}, \z^{1}_{2}, \dots, \z^{1}_{m_{1}}\}$. 
We have, for the case of $s$ sensors
\begin{align}
F[g_{1:s},u] &= \left( \prod_{j=1}^{s}{C_{j}(\Dotp{c_{j}}{g_{j}})} \right) \, 
	\pgf(\Dotp{r}{u\prod_{j=1}^{s}{\phi_{g_{j}}}}).
\end{align}

Differentiating the above expression with respect to the set $Z^{1}_{k+1}$ we get
\begin{align}
&\frac{\delta F}{\delta Z^{1}_{k+1}}[g_{1:s},u] \nonumber \\
&= \frac{\delta}{\delta Z^{1}_{k+1}} 
	\left\{ \left( \prod_{j=1}^{s}{C_{j}(\Dotp{c_{j}}{g_{j}})} \right) \, 
	\pgf(\Dotp{r}{u\prod_{j=1}^{s}{\phi_{g_{j}}}}) \right\} \\
&= \left( \prod_{j=2}^{s}{C_{j}(\Dotp{c_{j}}{g_{j}})} \right) \frac{\delta}{\delta Z^{1}_{k+1}} 
	\left\{ {C_{1}(\Dotp{c_{1}}{g_{1}})} \, \pgf(\Dotp{r}{u\prod_{j=1}^{s}{\phi_{g_{j}}}}) \right\}
\end{align}
since the differential $\frac{\delta}{\delta Z^{1}_{k+1}}$ only differentiates the 
variable $g_1$. If $I \subseteq \ll 1 , m_{1} \rr$ we can express 
$Y \subseteq Z^{1}_{k+1}$ as $Y = \{ \z^{1}_{i}: i \in I \}$ for some $I$. We also 
have $Z^{1}_{k+1} - Y = \{ \z^{1}_{i}: i \notin I \}$. Using the product rule for 
functional derivatives from~\eqref{eq:product_rule_2} we have
\begin{align}
&\frac{\delta}{\delta Z^{1}_{k+1}} \left\{ {C_{1}(\Dotp{c_{1}}{g_{1}})} \, 
	\pgf(\Dotp{r}{u\prod_{j=1}^{s}{\phi_{g_{j}}}}) \right\} \nonumber \\
&= \sum_{n=0}^{m_{1}} \sum_{\substack{I \subseteq \ll 1 , m_{1} \rr \\ |I| = n}}
	\frac{\delta}{\delta \{\z^{1}_{i}\}_{i \in I}} \pgf(\Dotp{r}{u\prod_{j=1}^{s}{\phi_{g_{j}}}})
	\frac{\delta}{\delta \{\z^{1}_{i}\}_{i \notin I}} {C_{1}(\Dotp{c_{1}}{g_{1}})}.
	\label{eq:rhs_1}
\end{align}
Now we consider the derivatives of each of the individual terms in the above expression.
By application of the chain rule for functional derivatives~\cite[Ch.~11]{mahler2007B}
\begin{align}
&\frac{\delta}{\delta \{\z^{1}_{i}\}_{i \in I}} \pgf(\Dotp{r}{u\prod_{j=1}^{s}{\phi_{g_{j}}}}) \nonumber \\
&= \pgf^{(n)}(\Dotp{r}{u\prod_{j=1}^{s}{\phi_{g_{j}}}}) \, 
	\prod_{i \in I}{\Dotp{r}{u\,p^{1}_{d}\,h_{1}(\z^{1}_{i})\,\prod_{j=2}^{s}{\phi_{g_{j}}}}}.
\end{align}
In the expression above and in later expressions we have used the convention that whenever $|I| = 0$, $\prod_{i \in I}() = 1$.
Applying the chain rule to the second derivative
\begin{align}	
\frac{\delta}{\delta \{\z^{1}_{i}\}_{i \notin I}} {C_{1}(\Dotp{c_{1}}{g_{1}})} 
&= C^{(m_{1}-n)}_{1}(\Dotp{c_{1}}{g_{1}}) \; \prod_{i \notin I}{c_{1}(\z^{1}_{i})} \\
&= C^{(m_1-n)}_{1}(\Dotp{c_{1}}{g_{1}}) \; \frac{\Gamma^{(1)}}{\displaystyle \prod_{i \in I}{c_{1}(\z^{1}_{i})}}.
\end{align}
As before we have used the convention $\prod_{i \in I}() = 1$ when $|I| = 0$.
Thus the right hand side of \eqref{eq:rhs_1} can be expressed as
\begin{align}
\Gamma^{(1)} &\, \sum_{n=0}^{m_{1}} \sum_{\substack{I \subseteq \ll 1 , m_{1} \rr \\ |I| = n}} 
	\Bigg\{ \pgf^{(n)}(\Dotp{r}{u\prod_{j=1}^{s}{\phi_{g_{j}}}}) \times \nonumber \\ 
& C^{(m_1-n)}_{1}(\Dotp{c_{1}}{g_{1}}) 
	\prod_{i \in I}{\frac{\Dotp{r}{u\,p^{1}_{d}\,h_{1}(\z^{1}_{i})\,\prod_{j=2}^{s}{\phi_{g_{j}}}}}
	{c_{1}(\z^{1}_{i})}} \Bigg\}.
\end{align}
In the double summation above, each set $I$ maps to a partition $P$ of the form
$P = \bigcup_{i \in I}{ \{ \z^{1}_{i} \} }$ in $\S^{(1)}$ and vice versa. Hence 
using result in equation~\eqref{eq:union_r1} of Appendix~\ref{app:recursive} we have
\begin{align}
&\frac{\delta F}{\delta Z^{1}_{k+1}}[g_{1:s},u] = \Gamma^{(1)} \sum_{P \in \S^{(1)}} \Bigg\{
	C^{(m_1-|P|_{1})}_{1}(\Dotp{c_{1}}{g_{1}}) \times \nonumber \\
& \left( \prod_{j=2}^{s}{C_{j}(\Dotp{c_{j}}{g_{j}})} \right)
	\pgf^{(|P|-1)}(\Dotp{r}{u\prod_{j=1}^{s}{\phi_{g_{j}}}}) \; 
	\prod_{W \in P}{\varphi_{W}[g_{1:s},u]} \Bigg\}
\end{align}	
Note that for the empty partition $P = \{V\}$, there are no elements of the form $W$ in $P$.
Hence in the above expression we use the convention $\prod_{W \in P}() = 1$ whenever $P = \{V\}$.
Further grouping of the terms gives the compact expression
\begin{align}
\frac{\delta F}{\delta Z^{1}_{k+1}}[g_{1:s},u] = \Gamma^{(1)} \sum_{P \in \S^{(1)}} \Psi_{P}^{(1)}[g_{1:s},u] \prod_{W \in P}{\varphi_{W}[g_{1:s},u]}.
\end{align}
Hence the result is established for the case $\l = 1$.
\newline
{\bf Mathematical induction: case $\l = b \geq 1$}

Now assuming that the result is true for some $\l = b \geq 1$, we establish 
that the result holds for $\l = b+1 \leq s$. 
Let $Z^{b+1}_{k+1} = \{ \z^{b+1}_{1}, \, \z^{b+1}_{2}, \dots, \z^{b+1}_{m_{b+1}}\}$.
We can write
\begin{align}
\frac{\delta F}{\delta Z^{1:b+1}_{k+1}}[g_{1:s},u] &= \frac{\delta}{\delta Z^{b+1}_{k+1}} 
	\left\{ \frac{\delta F}{\delta Z^{1:b}_{k+1}}[g_{1:s},u] \right\}.
\end{align}
Substituting the result for the case $\l = b$ we get
\begin{align}
&\frac{\delta F}{\delta Z^{1:b+1}_{k+1}}[g_{1:s},u] \nonumber \\
&= \frac{\delta}{\delta Z^{b+1}_{k+1}} \Bigg\{ \Gamma^{(b)} \sum_{P \in \S^{(b)}} 
	\Psi_{P}^{(b)}[g_{1:s},u] \prod_{W \in P}{\varphi_{W}[g_{1:s},u]} \Bigg\} \\
&= \Gamma^{(b)} \left( \prod_{j=b+2}^{s} {C_{j}(\Dotp{c_{j}}{g_{j}})} \right) \sum_{P \in \S^{(b)}}
	 \left( \prod_{j=1}^{b} {C^{(m_{j}-|P|_{j})}_{j}(\Dotp{c_{j}}{g_{j}})} \right) \nonumber \\
& \quad \times \frac{\delta}{\delta Z^{b+1}_{k+1}} \Bigg\{ C_{b+1}(\Dotp{c_{b+1}}{g_{b+1}}) 
	\pgf^{(|P|-1)}(\Dotp{r}{u \prod_{j=1}^{s}{\phi_{g_{j}}}}) \times \nonumber \\ 
& \qquad \qquad \prod_{W \in P}{\varphi_{W}[g_{1:s},u]}\Bigg\}. \label{eq:rhs_2}
\end{align}

Let $I_{1} \subseteq \ll 1 , m_{b+1} \rr$ and $I_{2} \subseteq \ll 1 , m_{b+1} \rr$ 
such that $I_{1} \cap I_{2} = \emptyset$. Then we can express $Y_{1} \subseteq Z^{b+1}_{k+1}$
and $Y_{2} \subseteq Z^{b+1}_{k+1}$ satisfying $Y_{1} \cap Y_{2} = \emptyset$ as
$Y_{1} = \{ \z^{b+1}_{i}: i \in I_{1} \}$ and $Y_{2} = \{ \z^{b+1}_{i}: i \in I_{2} \}$
respectively. Applying the product rule from~\eqref{eq:product_rule_3} to the 
expression above we have
\begin{align}
& \frac{\delta}{\delta Z^{b+1}_{k+1}} \Bigg\{ C_{b+1}(\Dotp{c_{b+1}}{g_{b+1}}) 
	\pgf^{(|P|-1)}(\Dotp{r}{u \prod_{j=1}^{s}{\phi_{g_{j}}}}) \times \nonumber \\ 
& \qquad \qquad \prod_{W \in P}{\varphi_{W}[g_{1:s},u]}\Bigg\} \nonumber \\
& \; = \sum_{n_{1}=0}^{m_{b+1}} \sum_{n_{2}=0}^{\textrm{min}(m_{b+1},|P|-1)}
	\sum_{\substack{I_{1} \subseteq \ll 1 , m_{b+1} \rr \\ |I_{1}| = n_{1}}}
	\sum_{\substack{I_{2} \subseteq \ll 1 , m_{b+1} \rr \\ |I_{2}| = n_{2}; \; I_{1} \cap I_{2} = \emptyset}} \nonumber \\
& \qquad \Bigg\{ \frac{\delta}{\delta \{\z^{b+1}_{i_{1}}\}_{i_{1} \in I_{1}}} 
	\pgf^{(|P|-1)}(\Dotp{r}{u \prod_{n=1}^{s}{\phi_{g_{n}}}}) \times \nonumber \\
&	\qquad \qquad \frac{\delta}{\delta \{\z^{b+1}_{i_{2}}\}_{i_{2} \in I_{2}}}
	\left( \prod_{W \in P}{\varphi_{W}[g_{1:s},u]} \right) \times \nonumber \\
& \qquad \qquad \frac{\delta}{\delta \{\z^{b+1}_{i}\}_{i \notin I_{1} \cup I_{2}}} 
	C_{b+1}(\Dotp{c_{b+1}}{g_{b+1}}) \Bigg\} .
\end{align}
The second summation above is restricted to the limit $\textrm{min}(m_{b+1},|P|-1)$ 
because the derivatives of $\prod_{W \in P}{\varphi_{W}[g_{1:s},u]}$ for $n_{2} > |P|-1$ 
are zero.
Now considering each of the individual derivatives above we have
\begin{align}
& \frac{\delta}{\delta \{\z^{b+1}_{i_{1}}\}_{i_{1} \in I_{1}}} 
	\pgf^{(|P|-1)}(\Dotp{r}{u \prod_{n=1}^{s}{\phi_{g_{n}}}}) \nonumber \\
&= \pgf^{(|P|+n_{1}-1)}(\Dotp{r}{u \prod_{n=1}^{s}{\phi_{g_{n}}}}) 
	\prod_{i_{1} \in I_{1}}{\varphi_{\{\z^{b+1}_{i_{1}}\}}[g_{1:s},u] \, c_{b+1}(\z^{b+1}_{i_{1}})}.
\end{align}
Denote $P = \{ W_{1}, W_{2}, \dots, W_{|P|-1}, V \}$ for notational convenience. 
Then we have
\begin{align}
\frac{\delta}{\delta \{\z^{b+1}_{i_{2}}\}_{i_{2} \in I_{2}}} &
	\left( \prod_{W \in P}{\varphi_{W}[g_{1:s},u]} \right) \nonumber \\
&= \sum_{\substack{J \subseteq \ll 1 , |P|-1 \rr \\ |J| = |I_{2}|}} \sum_{B \in \mathcal{B}(I_{2},J)} 
	\Bigg\{ \left( \prod_{j \notin J}{\varphi_{W_{j}}[g_{1:s},u]} \right) \times \nonumber \\
& \qquad \left( \prod_{i_{2} \in I_{2}}{\varphi_{W^{B}_{i_{2}}}
	[g_{1:s},u] \, c_{b+1}(\z^{b+1}_{i_{2}})} \right) \Bigg\}
\end{align}
where $\mathcal{B}(I_{2},J)$ is the collection of all possible matchings 
from set $I_{2}$ to set $J$ and we define the measurement subset 
$W^{B}_{i_{2}} \eqdef W_{B(i_{2})} \cup \z^{b+1}_{i_{2}}$.
Also
\begin{align}
& \frac{\delta}{\delta \{\z^{b+1}_{i}\}_{i \notin I_{1} \cup I_{2}}} 
	\left( C_{b+1}(\Dotp{c_{b+1}}{g_{b+1}}) \right) \nonumber \\
&= C^{(m_{b+1}-n_{1}-n_{2})}_{b+1}(\Dotp{c_{b+1}}{g_{b+1}}) 
	\prod_{i \notin I_{1} \cup I_{2}}{c_{b+1}(\z^{b+1}_{i})}.
\end{align}

Combining the three derivatives into the right hand side of 
expression \eqref{eq:rhs_2} we get
\newpage
\begin{align}
& \Gamma^{(b)} \left( \prod_{j=b+2}^{s} {C_{j}(\Dotp{c_{j}}{g_{j}})} \right)
	\left( \prod_{\z^{b+1} \in Z^{b+1}_{k+1}}{\!\!\!\! c_{b+1}(\z^{b+1})}\right) \nonumber \\
& \sum_{P \in \S^{(b)}} \sum_{n_{1}=0}^{m_{b+1}} \sum_{n_{2}=0}^{\textrm{min}(m_{b+1},|P|-1)}
	\!\!\!\!\!\! \sum_{\substack{I_{1} \subseteq \ll 1 , m_{b+1} \rr \\ |I_{1}| = n_{1}}}
	\sum_{\substack{I_{2} \subseteq \ll 1 , m_{b+1} \rr \\ |I_{2}| = n_{2}; \; I_{1} \cap I_{2} = \emptyset}}
  \sum_{\substack{J \subseteq \ll 1 , |P|-1 \rr \\ |J| = |I_{2}|}}  \nonumber \\ 
& \sum_{B \in \mathcal{B}(I_{2},J)} \Bigg\{ C^{(m_{b+1}-n_{1}-n_{2})}_{b+1}(\Dotp{c_{b+1}}{g_{b+1}})
	\left( \prod_{j \notin J}{\varphi_{W_{j}}[g_{1:s},u]} \right) \times \nonumber \\
& \quad \pgf^{(|P|+n_{1}-1)}(\Dotp{r}{u \prod_{n=1}^{s}{\phi_{g_{n}}}}) \,
	\left( \prod_{j=1}^{b} {C^{(m_{j}-|P|_{j})}_{j}(\Dotp{c_{j}}{g_{j}})} \right) \times \nonumber \\
& \quad \left( \prod_{i_{1} \in I_{1}}{\varphi_{\{\z^{b+1}_{i_{1}}\}}[g_{1:s},u]} \right)
	\left( \prod_{i_{2} \in I_{2}}{\varphi_{W^{B}_{i_{2}}}[g_{1:s},u]} \right) \Bigg\}.
\end{align}
Using result of Appendix~\ref{app:recursive} we can simplify the multiple
summation term and write
\begin{align}
&\frac{\delta F}{\delta Z^{1:b+1}_{k+1}}[g_{1:s},u] =
  \Gamma^{(b+1)} \!\!\!\! \sum_{P \in \S^{(b+1)}} \Bigg\{
	\left( \prod_{j=1}^{b+1} {C^{(m_{j}-|P|_{j})}_{j}(\Dotp{c_{j}}{g_{j}})} \right) \nonumber \\
& \qquad \left( \prod_{j=b+2}^{s} {C_{j}(\Dotp{c_{j}}{g_{j}})} \right) 
	\pgf^{(|P|+n_{1}-1)}(\Dotp{r}{u \prod_{n=1}^{s}{\phi_{g_{n}}}}) \times \nonumber \\
&	\qquad \left( \prod_{W \in P}{\varphi_{W}[g_{1:s},u]} \right) \Bigg\} \nonumber \\
&= \Gamma^{(b+1)} \sum_{P \in \S^{(b+1)}} \Psi_{P}^{(b+1)}[g_{1:s},u]
	\prod_{W \in P}{\varphi_{W}[g_{1:s},u]}.
\end{align}
Hence we have established the result stated in~\eqref{eq:induction_statement}
using the method of mathematical induction. We obtain the result of 
Lemma~\ref{lem:F_derivative} by substituting $\l = s$ in this result.

\end{proof}

\section{Proof of Theorem~\ref{thm:phd_update}}
\label{app:phd_update}

For brevity denote $\Psi_{P}[0,0,\dots,0,u] = \Psi_{P}[u]$ and 
$\varphi_{W}[0,0,\dots,0,u] = \varphi_{W}[u]$. Substituting $g_{j} \equiv 0$
for $j = 1,2,\dots,s$ in the result of Lemma~\ref{lem:F_derivative} we get
\begin{align}
\frac{\delta F}{\delta Z^{1:s}_{k+1}}[0,0,\dots,0,u] &= \Gamma 
	\sum_{P \in \S} \Psi_{P}[u] \prod_{W \in P}{\varphi_{W}[u]}.
	\label{eq:F_derv_0}
\end{align}
\newline
{\bf PHD update}

Differentiating equation (\ref{eq:F_derv_0}) with respect to set $\{\x\}$ we have
\begin{align}
& \frac{\delta F}{\delta \x \delta Z^{1:s}_{k+1}}[0,0,\dots,0,u] = 
\frac{\delta}{\delta \x} \left\{ \frac{\delta F} {\delta Z^{1:s}_{k+1}}[0,0,\dots,0,u] \right\} \\
&= \frac{\delta}{\delta \x} \left\{ \Gamma \sum_{P \in \S} \Psi_{P}[u] 
	\prod_{W \in P}{\varphi_{W}[u]} \right\} \\
&= \Gamma \sum_{P \in \S} \left( \prod_{j=1}^{s} {C^{(m_{j}-|P|_{j})}_{j}(0)} \right) \times \nonumber \\ 
&	\qquad \frac{\delta}{\delta \x} \left\{ \pgf^{(|P|-1)}(\Dotp{r}{u \prod_{j=1}^{s}{q^{j}_{d}}}) \, 
	\prod_{W \in P}{\varphi_{W}[u]} \right\}.
\end{align}
Applying the product rule for set derivatives from~\eqref{eq:product_rule_2}
\begin{align}
\frac{\delta}{\delta \x} &\left\{ \pgf^{(|P|-1)}(\Dotp{r}{u \prod_{j=1}^{s}{q^{j}_{d}}}) \, 
	\prod_{W \in P}{\varphi_{W}[u]} \right\} \nonumber \\
&= \frac{\delta}{\delta \x} \left\{ \pgf^{(|P|-1)}(\Dotp{r}{u \prod_{j=1}^{s}{q^{j}_{d}}}) \right\} \, 
	\prod_{W \in P}{\varphi_{W}[u]} + \nonumber \\
& \quad \pgf^{(|P|-1)}(\Dotp{r}{u \prod_{j=1}^{s}{q^{j}_{d}}}) \frac{\delta}{\delta \x} \left\{ 
	\prod_{W \in P}{\varphi_{W}[u]} \right\}.
\end{align}
Evaluating the individual derivatives above and substituting the constant
function $u(\x) \equiv 1$, we get
\begin{align}
&\frac{\delta}{\delta \x} \left\{ \pgf^{(|P|-1)}(\Dotp{r}{u \prod_{j=1}^{s}{q^{j}_{d}}}) \right\}_{u \equiv 1}
= \pgf^{(|P|)}(\gamma) \, r(\x) \prod_{j=1}^{s}{q^{j}_{d}(\x)} \\
&\frac{\delta}{\delta \x} \left\{ \prod_{W \in P}{\varphi_{W}[u]} \right\}_{u \equiv 1}
= \left( \prod_{W \in P}{d_{W}} \right) \left( \sum_{W \in P} r(\x) \rho_{W}(\x) \right) \label{eq:ddx_varphi}
\end{align}
where $\gamma$, $d_{W}$ and $\rho_{W}(\x)$ are defined in~\eqref{eq:gamma},
\eqref{eq:dW_cphd} and~\eqref{eq:rhoW_cphd} respectively.
We note that for the empty partition $P = \{V\}$ there are no elements of the form $W$ in $P$.
In this case the derivative in equation \eqref{eq:ddx_varphi} is zero since the quantity being 
differentiated is a constant equal to 1. To have a compact representation
of the update equations we use the convention $\sum_{W \in P} () = 0$ when $P = \{V\}$.
Hence we have
\begin{align}
& \frac{\delta F}{\delta \x \delta Z^{1:s}_{k+1}}[0,0,\dots,0,1] = 
	\Gamma \sum_{P \in \S} \left( \prod_{j=1}^{s} {C^{(m_{j}-|P|_{j})}_{j}(0)} \right) \times \nonumber \\
& \quad \Bigg\{ \pgf^{(|P|)}(\gamma) \, r(\x) \prod_{j=1}^{s}{q^{j}_{d}(\x)} 
	\left( \prod_{W \in P}{d_{W}} \right) + \nonumber \\
& \qquad \pgf^{(|P|-1)}(\gamma) \left( \prod_{W \in P}{d_{W}} \right)
	\left( \sum_{W \in P}{r(\x)	\rho_{W}(\x)} \right) \Bigg\} \\
&= \Gamma \sum_{P \in \S} \left( \kappa_{P} \pgf^{(|P|)} \prod_{W \in P}{d_{W}} \right) 
	\, \left(r(\x) \prod_{j=1}^{s}{q^{j}_{d}(\x)}\right) \nonumber \\
& \quad + \Gamma \sum_{P \in \S} \left( \kappa_{P} \pgf^{(|P|-1)} \prod_{W \in P}{d_{W}} \right) 
	\, \left(\sum_{W \in P}{r(\x) \, \rho_{W}(\x)}\right) \label{eq:phd_nume}
\end{align}
where $\kappa_{P}$ is defined in~\eqref{eq:kappa_P}.
Substituting $u(\x) \equiv 1$ in equation (\ref{eq:F_derv_0}) we have
\begin{align}
\frac{\delta F}{\delta Z^{1:s}_{k+1}}[0,0,\dots,0,1] &= \Gamma 
	\sum_{P \in \S} \kappa_{P} \pgf^{(|P|-1)} \prod_{W \in P}{d_{W}}. \label{eq:phd_deno}
\end{align}
Dividing~\eqref{eq:phd_nume} by~\eqref{eq:phd_deno} and using the definition 
of PHD from~\eqref{eq:phd_defn_gen},
we get 
\begin{align}
& D_{k+1|k+1}(\x) = \frac{\displaystyle \frac{\delta F}{\delta \x \delta Z^{1:s}_{k+1}}[0,0,\dots,0,1]}
	{\displaystyle \frac{\delta F}{\delta Z^{1:s}_{k+1}}[0,0,\dots,0,1]} \\
& \quad = r(\x) \left\{ \alpha_{0} \, \prod_{j=1}^{s}{q^{j}_{d}(\x)} + \sum_{P \in \S} 
	\alpha_{P} \, \left(\sum_{W \in P} \rho_{W}(\x) \right)	\right\}
\end{align}
where $\alpha_{0}$ and $\alpha_{P}$ are as given in~\eqref{eq:alpha_equation_0}
and~\eqref{eq:alpha_equation_P}.

{\bf Cardinality update}

We now derive the update equation for the posterior cardinality 
distribution. Using the expression for the posterior probability 
generating functional in~\eqref{eq:G_defn_gen} and the results 
of~\eqref{eq:F_derv_0} and~\eqref{eq:phd_deno} we have
\begin{align}
G_{k+1|k+1}[u] &= \frac{\displaystyle \sum_{P \in \S} \Psi_{P}[u] \prod_{W \in P}{\varphi_{W}[u]}}
	{\displaystyle \sum_{P \in \S} \kappa_{P} \pgf^{(|P|-1)} \prod_{W \in P}{d_{W}}}.
\end{align}
The probability generating function $\pgf_{k+1|k+1}(t)$ of the posterior 
cardinality distribution is obtained by substituting the constant 
function $u(\x) \equiv t$ in the expression for $G_{k+1|k+1}[u]$. 
Thus
\begin{align}
\pgf_{k+1|k+1}(t) 
&= \frac{\displaystyle \sum_{P \in \S} \Psi_{P}[t] \prod_{W \in P}{\varphi_{W}[t]}}
		{\displaystyle \sum_{P \in \S} \kappa_{P} \pgf^{(|P|-1)} \prod_{W \in P}{d_{W}}}.
\end{align}
For constant $t$ we have 
\begin{align}
\displaystyle \prod_{W \in P}{\varphi_{W}[t]} &= t^{|P|-1} \prod_{W \in P}{d_{W}} \\
\text{and } \Psi_{P}[t] &= \left( \prod_{j=1}^{s} {C^{(m_{j}-|P|_{j})}_{j}(0)} \right) \pgf^{(|P|-1)}(t \gamma).
\end{align}
Since $\pgf_{k+1|k+1}(t)$ is the PGF corresponding to the cardinality
distribution $p_{k+1|k+1}(n)$,
\begin{align}
&p_{k+1|k+1}(n) = \frac{1}{n!} \, \pgf^{(n)}_{k+1|k+1}(0) \\
&= \frac{1}{n!} \, \left\{ \frac{d^n}{d t^n} \frac{\displaystyle \sum_{P \in \S} 
	\Psi_{P}[t] t^{|P|-1} \prod_{W \in P}{d_{W}}}
	{\displaystyle \sum_{P \in \S} \kappa_{P} \pgf^{(|P|-1)} \prod_{W \in P}{d_{W}}} \right\}_{t = 0} \\
&= \frac{\displaystyle \sum_{P \in \S} \left( \prod_{j=1}^{s} 
	{C^{(m_{j}-|P|_{j})}_{j}(0)} \right) \, \prod_{W \in P}{d_{W}} 
	\left\{ \frac{d^n}{d t^n} t^{|P|-1} \, \pgf^{(|P|-1)}(t \gamma) \right\}_{t = 0}}
	{n! \, \displaystyle \sum_{P \in \S} \kappa_{P} \pgf^{(|P|-1)} \prod_{W \in P}{d_{W}}}.
\end{align}
Evaluating the derivative we get
\begin{align}
\bigg\{ \frac{d^n}{d t^n} & t^{|P|-1} \, \pgf^{(|P|-1)}(t \gamma) \bigg\}_{t=0} \nonumber \\
&= \begin{cases}
		0 & \text{if $n < |P|-1$} \\
		\displaystyle \frac{n!}{(n-|P|+1)!} \pgf^{(n)}(0) \, \gamma^{n-|P|+1} & \text{if $n \geq |P|-1$}.
	\end{cases}
\end{align}
We also have $\pgf^{(n)}(0) = n! \, p_{k+1|k}(n)$, hence
\begin{align}
& p_{k+1|k+1}(n) = p_{k+1|k}(n) \, \times \nonumber \\
& \quad \frac{\displaystyle \sum_{\substack{P \in \S \\ |P| \leq n+1}} \frac{n!}{(n-|P|+1)!} 
	\left( \prod_{j=1}^{s} {C^{(m_{j}-|P|_{j})}_{j}(0)} \right) \, \gamma^{n-|P|+1} \prod_{W \in P}{d_{W}}}
	{\displaystyle \sum_{P \in \S} \kappa_{P} \pgf^{(|P|-1)} \prod_{W \in P}{d_{W}}}.
\end{align}
We thus have
\begin{align}
\frac{p_{k+1|k+1}(n)}{p_{k+1|k}(n)} &= \frac{\displaystyle \sum_{\substack{P \in \S \\ |P| \leq n+1}} 
	\left(\kappa_{P} \frac{n!}{(n-|P|+1)!} \gamma^{n-|P|+1} \prod_{W \in P}{d_{W}}\right)}
	{\displaystyle \sum_{P \in \S} \kappa_{P} \pgf^{(|P|-1)} \prod_{W \in P}{d_{W}}}
\end{align}
where $\kappa_{P}$ is as defined in~\eqref{eq:kappa_P}.

\bibliographystyle{IEEEtran}
\bibliography{gcphd_bibtex}


\begin{IEEEbiography}[{\includegraphics[width=1in,height=1.25in,clip,keepaspectratio]{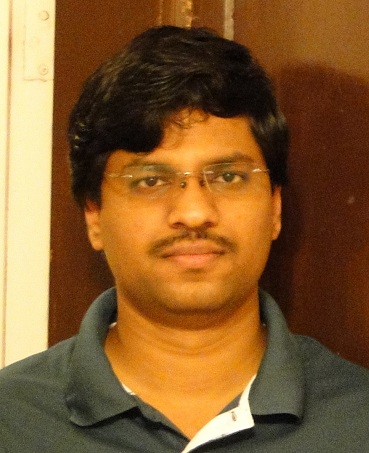}}]
{Santosh Nannuru} is currently a Postdoc in the Scripps Institution of Oceanography 
at University of California, San Diego. He obtained his doctorate in electrical engineering 
from McGill University, Canada in 2015. He received both his B.Tech and M.Tech Degrees 
in electrical engineering from Indian Institute of Technology, Bombay in 2009. He worked 
as a design engineer at iKoa Semiconductors for a year before starting his Ph.D. He is 
recipient of the McGill Engineering Doctoral Award (MEDA). His research interests are in 
sparse signal processing, Bayesian inference, Monte Carlo methods and random finite sets. 
\end{IEEEbiography}

\begin{IEEEbiography}[{\includegraphics[width=1in,height=1.25in,clip,keepaspectratio]{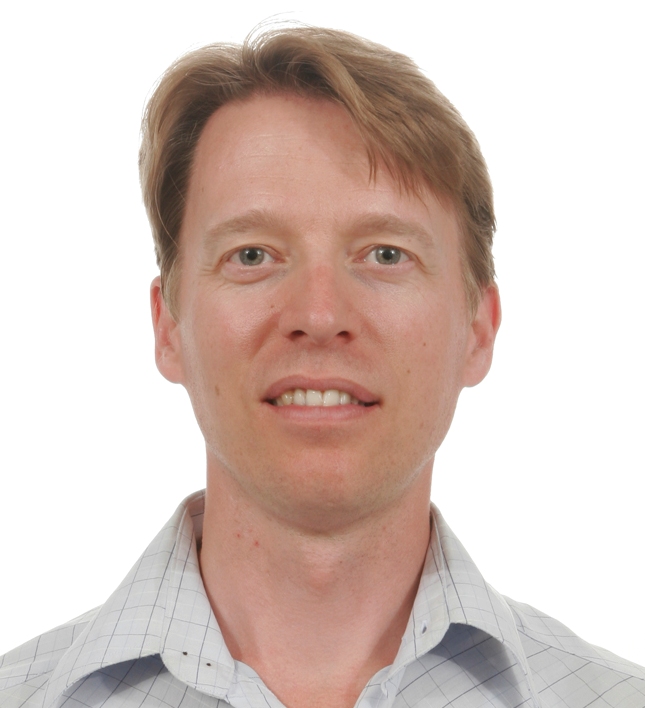}}]
{St\'{e}phane Blouin} is a professional engineer for 24 years. Dr. St\'{e}phane Blouin received a B.Sc. degree in mechanical engineering (Laval U. in Qu\'{e}bec city (QC), 1992), an M.Sc. degree in electrical engineering (Ecole Polytechnique in Montr\'{e}al (QC), 1995), and a Ph.D. degree in chemical engineering (Queen's U. in Kingston (ON), 2003). With more than 15 years of industrial experience, Dr.~Blouin held various R\&D positions in Canada, France and U.S.A. related to technology development and commercialization for automated processes, assembly lines, robotic systems, and process controllers. In 2010, he became a Defence Scientist at the Atlantic Research Centre of Defence R\&D Canada (DRDC). Dr.~Blouin is the first author of the paper which received the Best Paper Award at the 2015 International Conference on Sensor Networks (SENSORNETS). He currently holds adjunct professor positions at Dalhousie University (Halifax, Nova Scotia) and at Carleton University (Ottawa, Ontario). Dr.~Blouin has authored more than 50 scientific documents, holds 8 inventions and patents, and is the Canadian authority on many international projects. His current research interests include theoretical aspects of dynamic modeling, real-time monitoring, control, and optimization as well as experimental research applied to adaptive signal processing, sensors, distributed sensor networks, underwater networks, and intelligent unmanned systems.
\end{IEEEbiography}

\begin{IEEEbiography}[{\includegraphics[width=1in,height=1.25in,clip,keepaspectratio]{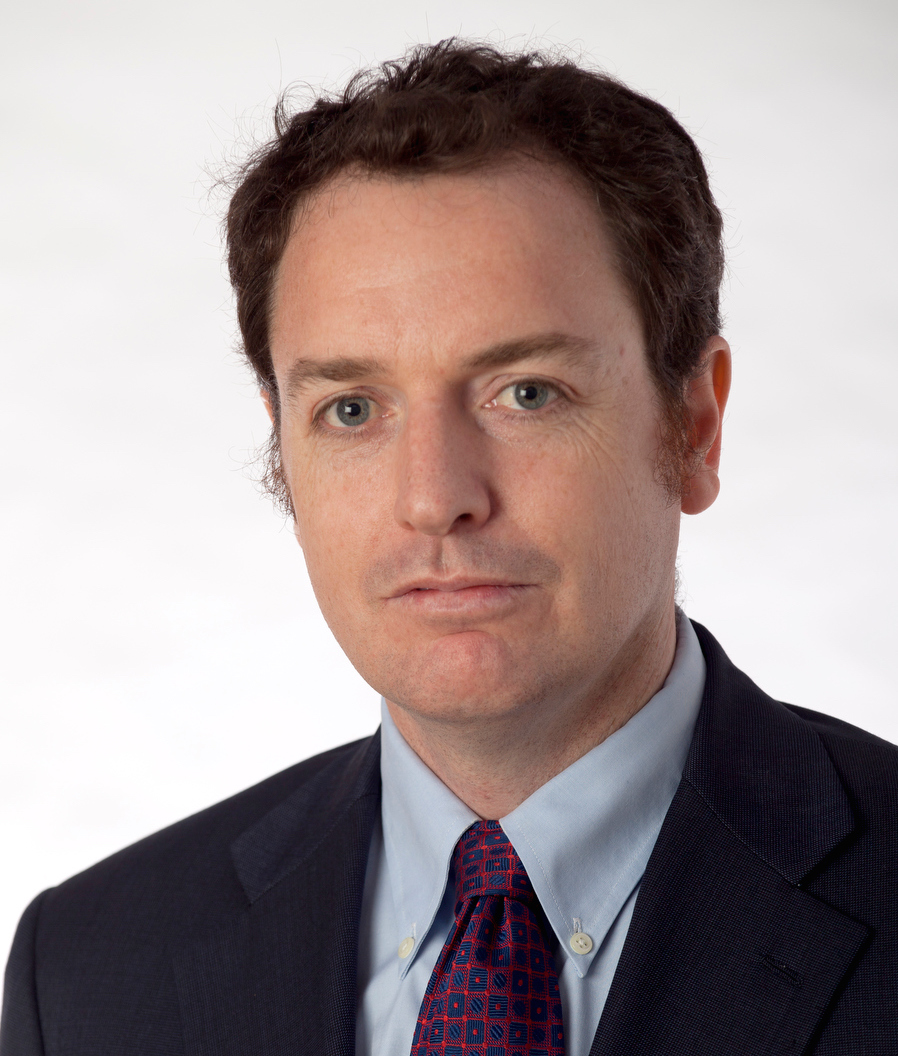}}]{Mark~J.~Coates}
  received the B.E. degree in computer systems engineering from the
  University of Adelaide, Australia, in 1995, and a Ph.D. degree in
  information engineering from the University of Cambridge, U.K., in
  1999. He joined McGill University (Montreal, Canada) in 2002, where
  he is currently an Associate Professor in the Department of
  Electrical and Computer Engineering. He was a research associate and
  lecturer at Rice University, Texas, from 1999-2001. In 2012-2013, he
  worked as a Senior Scientist at Winton Capital Management, Oxford,
  UK. He was an Associate Editor of IEEE Transactions on Signal
  Processing from 2007-2011 and a Senior Area Editor for
  IEEE Signal Processing Letters from 2012-2015. In 2006, his research team received
  the NSERC Synergy Award in recognition of their successful
  collaboration with Canadian industry, which has resulted in the
  licensing of software for anomaly detection and Video-on-Demand
  network optimization. Coates' research interests include
  communication and sensor networks, statistical signal processing,
  and Bayesian and Monte Carlo inference. His most influential and
  widely cited contributions have been on the topics of network
  tomography and distributed particle filtering. His contributions on
  the latter topic received awards at the International Conference on
  Information Fusion in 2008 and 2010. 
  \end{IEEEbiography}

\begin{IEEEbiography}
[{\includegraphics[width=1in,height=1.25in,clip,keepaspectratio]{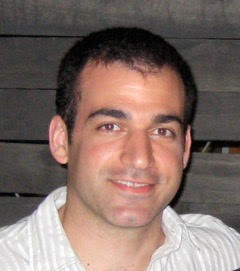}}]{Michael Rabbat} (S'02--M'07--SM'15) received the B.Sc.~degree from the University of Illinois, Urbana-Champaign, in 2001, the M.Sc.~degree from Rice University, Houston, TX, in 2003, and the Ph.D.~degree from the University of Wisconsin, Madison, in 2006, all in electrical engineering. He joined McGill University, Montr\'{e}al, QC, Canada, in 2007, and he is currently an Associate Professor. During the 2013--2014 academic year he held visiting positions at T\'{e}l\'{e}com Bretegne, Brest, France, the Inria Bretagne-Atlantique Reserch Centre, Rennes, France, and KTH Royal Institute of Technology, Stockholm, Sweden. He was a Visiting Researcher at Applied Signal Technology, Inc., Sunnyvale, USA, during the summer of 2003. Dr.~Rabbat co-authored the paper which received the Best Paper Award (Signal Processing and Information Theory Track) at the 2010 IEEE International Conference on Distributed Computing in Sensor Systems (DCOSS). He received an Honorable Mention for Outstanding Student Paper Award at the 2006 Conference on Neural Information Processing Systems (NIPS) and a Best Student Paper Award at the 2004 ACM/IEEE International Symposium on Information Processing in Sensor Networks (IPSN). He currently serves as Senior Area Editor for the \textsc{IEEE Signal Processing Letters} and as Associate Editor for \textsc{IEEE Transactions on Signal and Information Processing over Networks} and \textsc{IEEE Transactions on Control of Network Systems}. His research interests include distributed algorithms for optimization and inference, consensus algorithms, and network modelling and analysis, with applications in distributed sensor systems, large-scale machine learning, statistical signal processing, and social networks.
\end{IEEEbiography}

\end{document}